\documentclass[10pt,a4paper]{iopart}

\usepackage{amsmath}
\usepackage{stackengine} 
\usepackage{amssymb}
\usepackage{cite}
\usepackage{color}

\usepackage{tikz}
	\usetikzlibrary{backgrounds,calc,external}
	
\usepackage{pgfplots}
	\usepgfplotslibrary{fillbetween,groupplots}
	
\usepackage[pdftex,
	pdffitwindow	=true,
	pdfsubject 		={High-temperature plasma physics},
	pdfcreator 		={O. Linder},
	pdfproducer 	={O. Linder},
	pdftitle		={Self-consistent modeling of runaway electron generation in massive gas injection scenarios in ASDEX Upgrade},
	pdfauthor		={O Linder, E Fable, F Jenko, G Papp, G Pautasso, the ASDEX Upgrade Team, the EUROfusion MST1 Team},
	pdfkeywords		={runaway electrons, tokamaks, disruptions, massive material injection},
	]{hyperref}
	
\newcommand{\hs}{\hspace*{1pt}}

\newcommand\groupequation[2][10pt]{%
  \setbox0=\hbox{$\displaystyle#2$}%
  \stackengine{0pt}{\copy0}{%
    \makebox[\linewidth]{\hfill$\left.\rule{0pt}{\ht0}\right\}$\kern#1}}
    {O}{c}{F}{T}{L}
}

\newif\ifshowfigures
\showfigurestrue

\definecolor{ipp_blue}{rgb}{0.229739, 0.322361, 0.545706}
\definecolor{ipp_red}{HTML}{CF4D4A}
\definecolor{ipp_purple}{HTML}{A73C76}
\definecolor{ipp_green}{HTML}{96C144}
\definecolor{ipp_orange}{HTML}{CFA34A}	
	
\definecolor{set1_1}{HTML}{e41a1c}	
\definecolor{set1_2}{HTML}{377eb8}	
\definecolor{set1_3}{HTML}{4daf4a}	
\definecolor{set1_4}{HTML}{984ea3}	
\definecolor{set1_5}{HTML}{ff7f00}	
\definecolor{set1_6}{HTML}{ffff33}	
\definecolor{set1_7}{HTML}{a65628}	
\definecolor{set1_8}{HTML}{f781bf}	
\definecolor{set1_9}{HTML}{999999}	

\definecolor{set2_1}{HTML}{a6cee3}	
\definecolor{set2_2}{HTML}{1f78b4}	
\definecolor{set2_3}{HTML}{b2df8a}	
\definecolor{set2_4}{HTML}{33a02c}	
\definecolor{set2_5}{HTML}{fb9a99}	
\definecolor{set2_6}{HTML}{e31a1c}	
\definecolor{set2_7}{HTML}{fdbf6f}	
\definecolor{set2_8}{HTML}{ff7f00}	
\definecolor{set2_9}{HTML}{cab2d6}	
\definecolor{set2_10}{HTML}{6a3d9a}	

\definecolor{CET_0}{HTML}{000f5d}
\definecolor{CET_1}{HTML}{0d1575}
\definecolor{CET_2}{HTML}{291a88}
\definecolor{CET_3}{HTML}{4e1992}
\definecolor{CET_4}{HTML}{741392}
\definecolor{CET_5}{HTML}{940c8e}
\definecolor{CET_6}{HTML}{b00a88}
\definecolor{CET_7}{HTML}{ca1181}
\definecolor{CET_8}{HTML}{df2578}
\definecolor{CET_9}{HTML}{ee4069}
\definecolor{CET_10}{HTML}{f75c5a}
\definecolor{CET_11}{HTML}{f9784e}
\definecolor{CET_12}{HTML}{f99144}
\definecolor{CET_13}{HTML}{fba63b}
\definecolor{CET_14}{HTML}{fdbb38}
\definecolor{CET_15}{HTML}{fcd03c}
\definecolor{CET_16}{HTML}{f9e444}
\definecolor{CET_17}{HTML}{f5f94e}

\begin{document}

\title[Self-consistent modeling of RE generation in MGI scenarios in AUG]{Self-consistent modeling of runaway electron generation in massive gas injection scenarios in ASDEX Upgrade}

\author{	O Linder$^1$, 
			E Fable$^1$, 
			F Jenko$^1$,
			G Papp$^1$,
			G Pautasso$^1$,
			the ASDEX Upgrade Team\footnote{See author list of H. Meyer et al. 2019 Nucl. Fusion \textbf{59} \href{https://doi.org/10.1088/1741-4326/ab18b8}{112014}}
			and the EUROfusion MST1 Team\footnote{See author list of B. Labit et al. 2019 Nucl. Fusion \textbf{59} \href{https://doi.org/10.1088/1741-4326/ab2211}{086020}}}
			
\address{	$^1$ Max-Planck-Institut f\"ur Plasmaphysik, Boltzmannstr. 2, 85748 Garching, Germany}

\ead{\href{mailto:oliver.linder@ipp.mpg.de}{Oliver.Linder@ipp.mpg.de}}

\begin{abstract}%
\addcontentsline{toc}{section}{Abstract}
We present the first successful simulation of a induced disruption in ASDEX Upgrade from massive material injection (MMI) up to established runaway electron (RE) beam, thus covering pre-thermal quench, thermal quench and current quench (CQ) of the discharge. For future high-current fusion devices such as ITER, the successful suppression of REs through MMI is of critical importance to ensure the structural integrity of the vessel. To computationally study the interplay between MMI, background plasma response, and RE generation, a toolkit based on the 1.5D transport code coupling \texttt{ASTRA-STRAHL} is developed. Electron runaway is described by state-of-the-art reduced kinetic models in the presence of partially ionized impurities. Applied to argon MMI in ASDEX Upgrade discharge \#33108, key plasma parameters measured experimentally, such as temporal evolution of the line averaged electron density, plasma current decay rate and post-CQ RE current, are well reproduced by the simulation presented. Impurity ions are transported into the central plasma by the combined effect of neoclassical processes and additional effects prescribed inside the $q=2$ rational surface to explain experimental time scales. Thus, a thermal collapse is induced through strong impurity radiation, giving rise to a substantial RE population as observed experimentally.
\end{abstract}

\submitto{Nucl. Fusion}

\ioptwocol

\section{Introduction}
In the presence of sufficiently strong electric fields, plasma electrons surmount the collective collisional drag and experience net acceleration up to relativistic energies; a process referred to as electron runaway. Eventually, pitch-angle scattering and radiative losses due to Bremsstrahlung and synchrotron radiation hinder further gain of energy in the relativistic regime. These so called runaway electrons (RE) are created not only naturally in astrophysical or atmospheric plasmas, but also in human-made high-temperature laboratory plasmas. 

In current carrying fusion devices, electric fields sufficiently strong for the high-energy tail of the electron population to run away arise during the loss of both magnetic confinement and thermal energy, referred to as a disruption, following the law of induction as the plasma current and, more importantly, the magnetic field generated by the same collapses in the cold post-disruption plasma. Runaway electrons created in the process can potentially damage plasma facing components (PFC) severely upon impact by depositing both kinetic energy $W_\mathrm{kin} \propto \left\langle p_\mathrm{RE}^2\right\rangle I_\mathrm{RE}$ and magnetic energy $W_\mathrm{mag} \propto I_\mathrm{RE}^2$, with $p_\mathrm{RE}$ being the RE momentum and $I_\mathrm{RE}$ the RE carried current. Significant RE-induced damage of PCFs has been reported, e.g., at JET\cite{Reux15,Matthews16}, where RE currents close to 1~MA caused melting of substantial amounts of Be at the inner limiter of the ITER-like wall.

In ITER and other future high-current devices, REs pose an even greater threat to the structural integrity of the plasma vessel\cite{Hender07}. A large fraction of the pre-disruptive current $I_\mathrm{p}$ of up to 16~MA may be carried by suprathermal electrons post-disruption due to long exponentiation of a post-disruption RE seed, giving rise to runaway currents in excess of 10~MA. Simultaneously, the RE stored magnetic energy increases significantly compared to present-day devices, given the quadratic scaling with the current carried, $W_\mathrm{mag} \propto I_\mathrm{RE}^2$. Under these conditions, the PFCs of high-current devices may be damaged severely by deconfined REs.

To protect the plasma vessel of ITER during a disruption, massive material injection (MMI) into the plasma center on a ms-timescale is envisioned, thus evenly distributing the plasma stored thermal energy across the first wall through radiation. In the cold post-injection plasma, the noticeably increased density of plasma electrons enhances the collisional drag experienced by the electron population and prevents the formation of a considerable RE beam under ideal conditions.

Currently, the feasibility of MMI for RE suppression is being investigated in dedicated experiments across multiple machines using massive gas injection (MGI), e.g. at ASDEX Upgrade (AUG) \cite{Pautasso17} and TCV \cite{Coda19}, or shattered pellet injection (SPI), e.g. at DIII-D \cite{Commaux10}. Yet, extrapolation from present-day small and large machines to ITER is ambitious, given the exponential sensitivity of RE generation on the pre-disruptive plasma current. At the same time, elaborate models describing RE generation are being derived theoretically \cite{Hesslow18,Hesslow18_2,Hesslow19}. 

To complement experimental and theoretical studies of electron runaway in tokamak disruptions following MGI, the coupled 1.5D transport code toolkit \texttt{ASTRA-STRAHL} is presented in this work, allowing self-consistent simulations of the interactions between background plasma, impurity species and runaway electrons. The model is then applied for the simulation of argon (Ar) MGI in AUG discharge \#33108. Note, that simulations of MGI for disruption mitigation in ITER are presented in \cite{Konovalov14}. A description of the computational model employed in this work is provided in section \ref{sec:model_description}, covering i.a. the treatment of the impurity species and the theoretical models for RE generation incorporated into the tool. Experimental aspects of AUG discharge \#33108 are discussed in section \ref{sec:experimental_scenario}. Simulations of Ar MGI with \texttt{ASTRA-STRAHL} for this discharge are presented in section \ref{sec:simulations}. Final conclusions are drawn in section \ref{sec:conclusions}.

\section{Model description}
\label{sec:model_description}
The interaction between the tokamak background deuterium plasma, injected material and REs in artificially induced disruptions in AUG MGI scenarios is studied by transport modeling of particles and heat with the 1.5D transport code \texttt{ASTRA} \cite{Fable13} coupled to the impurity transport and radiation code \texttt{STRAHL} \cite{Dux99}. This toolset was previously used i.a. for the study of the pre-thermal quench (TQ) of AUG MGI experiments \cite{Fable16}, but is enhanced to simulate MGI up to the RE plateau phase, as outlined in section~\ref{sec:STRAHL}ff. An overview of the tool structure is shown in figure~\ref{fig:model_scheme}; the details explained in the following.

\subsection{Background plasma evolution}
\label{sec:background_plasma_evolution}
The evolution of the background plasma is performed by \texttt{ASTRA} through evaluation of the macroscopic transport equation
	\begin{align}
		\label{eq:astra_transport_equation}
		\frac{\partial Y}{\partial t} = \frac{1}{V'} \frac{\partial}{\partial \rho} \left( V' \left\langle \left( \Delta \rho \right)^2 \right\rangle \left\{ D \hspace*{1pt} \frac{\partial Y}{\partial \rho} - v \hspace*{1pt} Y\right\} \right) + \sum_j S_j
	\end{align}
for any of the evolved fluid quantities $Y(\rho, t)$ in the presence of diffusion $D$, advection $v$ and sources/sinks $S_j$. Here, $\rho$ is the toroidal flux-surface label. Hence, the quantities $\Delta \rho$ and $V' = \partial V/\partial \rho$ describe the magnetic geometry, where $V$ is the volume of a flux-surface. In this work, the transport equations for electron temperature $T_{\rm e}$, ion temperature $T_{\rm i}$, poloidal magnetic flux $\Psi$ and RE density $n_\mathrm{RE}$ are solved. The electron density $n_{\rm e}$ is obtained through quasi-neutrality from the ion densities $n_k(t)$ of each impurity species $k$ evolved by \texttt{STRAHL}; i.e. $n_{\rm e}(t) = n_{\rm D} + \sum_k \left\langle Z_k \right\rangle \hspace*{1pt} n_{k}(t)$, with average charge $\left\langle Z_k \right\rangle$. The deuterium population $n_{\rm D}$ is set to remain at the level prior to MGI. The initial profiles of the quantities being evolved are taken from experimental reconstruction. In the case of the initial profile of the poloidal magnetic flux, the reconstructed safety factor $q$ is used.

For the evolution of the electron temperature, the impurity radiation $P_{\rm rad}$ is used as a sink term. A global heat diffusion coefficient of the order of the power balance, i.e. $\chi_{\rm e} = 1~\mathrm{m}^2~\mathrm{s}^{-1}$, is applied to prevent the formation of strongly localized hot plasma beamlets during simulations, as observed e.g. in \cite{Putvinski97,Feher11}. Apart from this, the overall $T_{\rm e}$-evolution is unaffected by a reasonable choice of $\chi_{\rm e}$ in these scenarios, since radiated losses far outweigh conductive ones \cite{Feher11}. The temperature evolution of electrons and ions is coupled through heat exchange, described by $P_{\rm ei} = 3 (T_{\rm e} - T_{\rm i})n_{\rm e} m_{\rm e}/m_{\rm i} \tau_{\rm e}$ \cite{Fable13} ($\tau_{\rm e}$ being the electron collision time).

\hypertarget{model:equilibrium}{The} transport calculations for aforementioned quantities are performed in a realistic magnetic geometry obtained from the equilibrium solver \texttt{SPIDER} \cite{Ivanov05} or from a simple 3-moment solver, built into \texttt{ASTRA}. Applying the latter solver, the radial $r$ and vertical $z$ coordinates of flux-surfaces inside a tokamak of major radius $R_0$ are parametrized by \cite{Fable13}
	\begin{equation}
	\groupequation{
	\begin{split}
		r(a,\theta) &= R_0 + \Delta(a) + a \left\{ \cos \theta - \delta(a)\hs \sin^2 \theta \right\} ~, \\
		z(a,\theta) &= a\hs \kappa(a)\hs \sin \theta ~,
	\end{split}
	}
	\end{equation}
with mid-plane minor radius $a$ of a flux-surface and poloidal angle $\theta$. The Shafranov shift is denoted by $\Delta(a)$, the triangularity by $\delta(a)$ and elongation by $\kappa(a)$. Since AUG MGI experiments are typically performed in circular discharges to ensure vertical stability of the plasma during disruption, application of the 3-moment solver is sufficient for this study.

It should be noted, that the impact of MHD phenomena on the magnetic equilibrium is not included in the simulations. In particular, the occurrence of a spike of the total plasma current observed in Ar MGI experiments at AUG when the cold gas front reaches the $q=2$ surface is not taken into account. However, the impact of this MHD phenomenon on the plasma species may be mimicked by application of increased transport coefficients inside the $q=2$ surface. An estimate of the required magnitude can be obtained through comparison with experimental observations of kinetic profiles, as will be described in section~\ref{sec:impurity_mixing}.

\ifshowfigures %
  \tikzsetnextfilename{00_model_plot}%
  \begin{figure}[bt!]
	\begin{small}
	
	\tikzset{external/export next=false}
	
	\begin{tikzpicture}
		\tikzstyle{every node}=[line width=.1mm, rounded corners=.25mm]
		
		\node[	draw=set2_2, fill=set2_2!10, inner ysep=.05cm, outer sep=0,
				text width=0.4*\linewidth]
			(ASTRA) at (0.,0.)
		 	{%
		 		\vspace*{-.25cm}
		 		\begin{center}\begin{tikzpicture}
					\node[	draw=set2_2, fill=set2_2!5, inner sep=.1cm, 
							inner ysep=.05cm, outer sep=0,
							text width=0.95\linewidth] 
						(box) at (0,0) 
						{\rule{0pt}{3ex}\\\vspace*{-.35cm} plasma evolution};
						
					\node[	text=white, anchor=west, fill=set2_2,
							inner sep=.1cm,outer sep=.2cm,
							text width=.7\linewidth] 
						(box_label) at (box.north west) 
						{\hyperref[sec:background_plasma_evolution]{Core routines}};
				\end{tikzpicture}\vspace*{.1cm} 
				\begin{tikzpicture}		
					\node[	draw=set2_2, fill=set2_2!5, inner sep=.1cm, 
							inner ysep=.05cm, outer sep=0,
							text width=0.95\linewidth] 
						(box) at (0,0) 
						{\rule{0pt}{3ex}\\\vspace*{-.35cm} equilibrium};
						
					\node[	text=white, anchor=west, fill=set2_2,
							inner sep=.1cm,outer sep=.2cm,
							text width=.7\linewidth] 
						(box_label) at (box.north west) 
						{\hyperlink{model:equilibrium}{3M solver}};
				\end{tikzpicture}\vspace*{.1cm} 
				\begin{tikzpicture}
					\node[	draw=set2_2, fill=set2_2!5, inner sep=.1cm, 
							inner ysep=.05cm, outer sep=0,
							text width=0.95\linewidth] 
						(box) at (0,0) 
						{\rule{0pt}{3ex}\\\vspace*{-.35cm} RE sources};
						
					\node[	text=white, anchor=west, fill=set2_2,
							inner sep=.1cm,outer sep=.2cm,
							text width=.7\linewidth] 
						(box_label) at (box.north west) 
						{\hyperlink{model:REGIA}{\texttt{\textbf{REGIA}}}};
				\end{tikzpicture}\end{center}
		 	};

		\node[	text=white, anchor=west, fill=set2_2,
				inner sep=.1cm,outer sep=.2cm] 
			(three) at (ASTRA.north west) 
			{\texttt{\hyperref[sec:background_plasma_evolution]{\textbf{ASTRA}}}};
		
		\node[	draw=set2_2, fill=set2_2!10, inner ysep=.05cm, outer sep=0,
				text width=.4*\linewidth]
			(STRAHL) at (.55\linewidth,0.) 
			{%
				\centering
				\rule{0pt}{3ex}\\\vspace*{-.35cm} impurity evolution  \\ \vspace*{-.5cm}
				\begin{center}\begin{tikzpicture}
					\node[	draw=set2_2, fill=set2_2!5, inner sep=.1cm, 
							inner ysep=.05cm, outer sep=0,
							text width=0.95\linewidth] 
						(NEOART) at (0,0) 
						{\rule{0pt}{3ex}\\\vspace*{-.35cm} neoclassical transport};
						
					\node[	text=white, anchor=west, fill=set2_2,
							inner sep=.1cm,outer sep=.2cm,
							text width=.7\linewidth] 
						(NEOART_label) at (NEOART.north west) 
						{\hyperlink{model:NEOART}{\texttt{\textbf{NEOART}}}};
						
				\end{tikzpicture}\end{center}
			};
			
		\node[	text=white, anchor=west, fill=set2_2,
				inner sep=.1cm,outer sep=.2cm] 
				(STRAHL_label) at (STRAHL.north west) 
				{\texttt{\hyperref[sec:STRAHL]{\textbf{STRAHL}}}};

			\draw[-latex] 
				(ASTRA.north east) to[bend left=30] node[fill=white] 
					{\parbox{.15\linewidth}{$n_{\rm e}, T_{\rm e}, T_{\rm i},$\\geometry}}
				(STRAHL.north);
			\draw[-latex] 
				(STRAHL.south) to[bend left=30] node[fill=white!0.1]
					{\parbox{.15\linewidth}{$n_{k}, P_{{\rm rad},k},$\\$\left\langle Z_{k} \right\rangle, Z_{\rm eff}$}}
				(ASTRA.south east)
				;

	\end{tikzpicture}
	\end{small}
	\caption[Scheme of model \texttt{ASTRA-STRAHL}]{%
		\label{fig:model_scheme}%
		Overview of the model employed in this work. The transport code \texttt{ASTRA} evolves the main plasma in a magnetic equilibrium obtained by the built-in 3-moment solver. Kinetic profiles are passed to \texttt{STRAHL} to calculate the evolution of impurity species under consideration of neoclassical transport coefficients obtained through \texttt{NEOART} inside \texttt{STRAHL}. Impurity densities, radiation and average charge of each species $k$, as well as the plasma effective charge are passed back to \texttt{ASTRA}. In this environment, RE sources are calculated by \texttt{REGIA} and the runaway population is evolved by \texttt{ASTRA}.
	}
\end{figure}
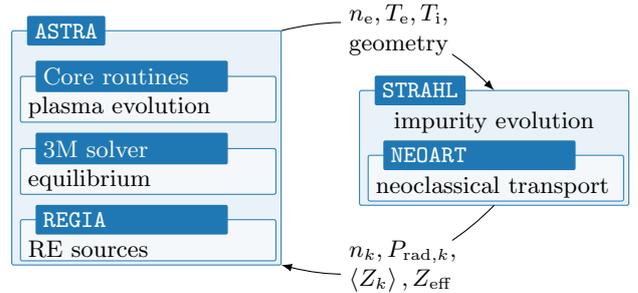%
 \fi

\subsection{Impurity transport modeling}
\label{sec:STRAHL}
The evolution of impurity species is calculated for each individual charge state by \texttt{STRAHL} from transport equations as used in \texttt{ASTRA} (see equation~\eqref{eq:astra_transport_equation}). The computational grid is provided by \texttt{ASTRA} to simplify the exchange of (kinetic) profiles between both codes. Additionally, electron density and temperature profiles are passed to \texttt{STRAHL} to evaluate reaction rates of atomic processes for a given impurity species. These electron-atom interactions, most importantly ionization and recombination, are described through tabulated rate coefficients from \texttt{ADAS}\cite{Summers04}. During evaluation of the impurity evolution, reaction rates obtained are prescribed as sources/sinks $S_j$. For each impurity species $k$, charge state $i$ resolved impurity densities $n_{k,i}$, total impurity radiation $P_{\mathrm{rad},k}$ and average impurity ion charge $\left\langle Z_k \right\rangle = \sum_{i\geq 1} Z_{k,i} \hs n_{k,i} /\sum_{i\geq 1} n_{k,i}$ are passed to \texttt{ASTRA} and are applied as described in section \ref{sec:background_plasma_evolution}. Additionally, the effective charge $Z_{\rm eff} = \{ n_{\rm D} + \sum_{k,i} Z_{k,i}^2 \hs n_{k,i} \}/ n_{\rm e}$ of the plasma under consideration of all plasma ions is returned.

\hypertarget{model:NEOART}{The} diffusive and convective impurity transport coefficients $D$ and $v$ are calculated in the case of neoclassical transport by the code \texttt{NEOART}\cite{Peeters00}. Geometric quantities for the calculation are provided by \texttt{ASTRA}. Increased simulation performance is achieved by parallelizing calls to \texttt{NEOART} for individual grid points with \texttt{openMP} and by applying \texttt{STRAHL}-specific optimization. Further transport mechanisms can be considered through externally provided transport coefficients. In this work, an additional global impurity diffusion with $D = 1~\mathrm{m}^2~\mathrm{s}^{-1}$ is applied; giving $D/\chi_e \sim 1$ as observed in gyrokinetic studies of impurity transport \cite{Angioni17}.

To enable the treatment of strongly advective particle transport, where the (local) P\'eclet number $\mu = v\Delta \rho/ D \to \infty$, the 1$^{\rm st}$ order central finite difference scheme used for the spatial discretization of the \texttt{STRAHL} transport equation is replaced with a finite volume discretization. The order of the scheme is kept to allow for continued application of the fast tridiagonal matrix algorithm during evaluation of the matrix equations. The numerical scheme used for discretization is chosen depending on the parameter $K = \max \left(0, 1 - 2/|\mu|\right) \cdot {\rm sign}(\mu)$\cite{Hundsdorfer03}. In diffusion dominated cases ($K \to 0$), a central scheme is applied, whereas in advection dominated cases ($K \to \infty$) upwinding is used. This approach prevents introducing numerical oscillations in the case $|\mu| \to \infty$, while simultaneously preventing application of an inaccurate scheme for $|\mu| \to 0$. As a result, impurity transport simulations with arbitrary P\'eclet number are enabled in \texttt{STRAHL}. As additional benefit of the finite volume scheme, particle conservation is ensured. Numerical details of the scheme are outlined in \ref{sec:appendix_fvm_STRAHL}.

\subsection{Neutral particle modeling}
Neutral impurities are treated similarly in \texttt{STRAHL} as their ionized counterparts. However, to distinguish between recombined impurity ions and externally deposited neutrals with different distribution functions $f(\rho, \mathbf{v}, t)$, multiple neutral impurity populations are employed within \texttt{STRAHL}. External neutral particles can be initialized at the start of the simulation with an arbitrary density profile or may be deposited gradually at an arbitrary location within the simulation domain throughout the course of the simulation. Once deposited, external neutrals are assumed to propagate inwards with thermal velocity $v_{\rm th} = \sqrt{T/m}$ until being ionized eventually. Depending on the temperature evolution during the simulation, ions may recombine to neutrals.

For the simulation of MGI in this work, the source of neutrals is set just outside the last closed flux surface. Consequently, the propagation from gas valve to the plasma edge is not modeled. Instead, experimental experience is used to estimate the neutral flow rate and flight time from gas valve to plasma boundary. 

\subsection{Runaway electron modeling}
\label{sec:RE_modeling}
The density of runaway electrons $n_{\rm RE}$ is treated in \texttt{ASTRA} as a separate species, following the same transport equations as described by equation~\eqref{eq:astra_transport_equation}. Source terms $S_j$ for RE generation considered in this study include the Dreicer mechanism and the avalanche mechanism. The former process describes small-angle momentum space diffusion of thermal electrons beyond the critical momentum for electron runaway (primary generation), whereas the latter one covers large-angle knock-on collisions of existing REs with the thermal bulk, thus creating secondary REs. Additional mechanisms for primary generation are not considered in this work. In AUG's non-nuclear environment, tritium $\beta^{-}$-decay and Compton scattering of high energy photons originating from activated wall material do not pose a relevant source of REs. Rapid cooling of the plasma in AUG may increase RE formation, as the high-energy tail of the electron distribution equilibrates slower and may therefore exceed the critical energy for runaway under these conditions. Still, this mechanism, referred to as hot-tail generation \cite{Chiu98,Harvey00}, is not taken into account given the lack of suitable numerical models. However, as finite magnetic flux is converted into REs during disruptions, it is expected that Dreicer and avalanche mechanism compensate the absence of the hot-tail source, resulting in similar amounts of REs \cite{Papp15}.

Both the Dreicer and the avalanche RE sources considered can be described by widely used equations. An analytic expression for Dreicer generation in the presence of an electric field $E_\parallel$ (here, parallel to the magnetic field lines) was derived by Connor \& Hastie \cite{Connor75}, 
	\begin{align}
	\label{eq:Dreicer_growth_rate_CH}
		S_{\rm D} &= k_\mathrm{D}\hspace*{1pt} n_\mathrm{e}\hspace*{1pt}  \nu_\mathrm{e} \left(\frac{E_\parallel}{E_\mathrm{D}}\right)^{\hspace*{-3pt}-h} \exp \left( - \lambda \frac{E_\mathrm{D}}{E_\parallel} - \eta \sqrt{\frac{E_\mathrm{D}}{E_\parallel}}\right) ~, 
	\end{align}
where the unknown constant $k_\mathrm{D}$ is of order unity and is therefore taken as $k_\mathrm{D} = 1$. The quantity $\nu_\mathrm{e}$ denotes the thermal electron-electron collision frequency. Setting $\varepsilon = E_\parallel/E_\mathrm{c}$, the remaining numerical factors are given by
	\begin{equation}
	\groupequation{
	\begin{split}
			h &= \frac{Z_\mathrm{eff}+1}{16} \left\{ 1 + 2 \frac{ \varepsilon -2 }{\varepsilon - 1 } \sqrt{\frac{\varepsilon}{\varepsilon - 1}} \right\} + \frac{2}{\varepsilon - 1} ~, \\
			\lambda &= 2 \varepsilon \left\{ \varepsilon - \frac{1}{2}  - \sqrt{\varepsilon\left( \varepsilon -1 \right)}\right\} ~, \\
			\eta &= \varepsilon \sqrt{\frac{Z_\mathrm{eff}+1}{4\left( \varepsilon -1 \right)}} \left\{ \frac{\pi}{2} - \arcsin\left( 1 - \frac{2}{\varepsilon} \right) \right\} ~.
	\end{split}
	}
	\end{equation}
%

The characteristic fields, being the Dreicer field $E_\mathrm{D} = n_\mathrm{e}\hs e^3 \hs \ln \Lambda/4 \pi  \varepsilon_0^2 \hs m_\mathrm{e} \hs v_\mathrm{th}^2$ (in SI units) and the critical field $E_\mathrm{c} = E_\mathrm{D} \hs v_\mathrm{th}^2/c^2$, describe runaway of electrons at thermal and relativistic velocities, respectively. The critical field $E_\mathrm{c}$ constitutes the minimum field for net acceleration of electrons. For $E_\parallel > E_\mathrm{c}$, more energy is gained from the electric field during one collision time $\nu_\mathrm{e}^{-1}$ than transferred to low energy electrons. Consequently, the RE growth rate of equation~\eqref{eq:Dreicer_growth_rate_CH} vanishes for $E_\parallel/E_\mathrm{c} \to 1^+$, as the exponent $h \to -\infty$ and $E_\parallel \ll E_\mathrm{D}$ under these conditions.

An expression for generation of runaway electrons due to the avalanche mechanism was initially obtained by Rosenbluth \& Putvinski \cite{Rosenbluth97} 
	\begin{equation}
	\label{eq:Avalanche_growth_rate_RP}
	\groupequation{
	\begin{split}
		S_{\rm av} &= \hs n_\mathrm{RE} \frac{e^2}{m_\mathrm{e} \hs \Lambda_{\rm c} } \sqrt{\frac{\pi\hs \gamma}{3\left(Z_\mathrm{eff} + 5\right)}} \left\{ E_\parallel - E_\mathrm{c} \right\} \\
		&\times \left( 1 - \varepsilon + \frac{\frac{4\hs \pi}{3\hs\gamma} \left( Z_\mathrm{eff} + 1\right)^{2}}{\left( Z_\mathrm{eff} + 5\right)\left\{ \varepsilon^2 + \frac{4}{\gamma^2} -1 \right\}}\right)^{-\frac{1}{2}} ~, \\
		\gamma &= \left( 1 + 1.46\hs \sqrt{\epsilon} + 1.72\hs \epsilon \right)^{-1} ~,
	\end{split}
	}
	\end{equation}
where $\epsilon$ denotes the inverse tokamak aspect ratio. The quantity $\ln \Lambda_{\rm c}$ describes the Coulomb logarithm for relativistic electron-electron collisions. Although this expression is accurate for large fields, the growth rate in the vicinity of the critical field is underestimated \cite{Rosenbluth97}.

The two models discussed are valid only for plasmas consisting of fully ionized species. Yet following MGI, already low-$Z$ to medium-$Z$ impurities are only partially ionized, particularly in a cold post-TQ plasma. Under these conditions, Rosenbluth and Putvinski \cite{Rosenbluth97} suggested to generalize the expression of the critical electric field $E_{\rm c}$ by including half the bound electrons. An improved description is achieved by taking the impact of partially ionized impurities on RE dynamics directly into account. Increased friction, pitch-angle scattering and radiation losses enhance significantly both the critical field $E_{\rm c}$ and the avalanche growth rate $S_{\rm av}$ \cite{Hesslow18,Hesslow18_2,Hesslow19} beyond the the classical formulae of reference~\cite{Rosenbluth97}. In the presence of partially ionized impurities, the avalanche growth rate can be expressed as \cite{Hesslow19}
	\begin{align}
	\label{eq:Avalanche_growth_rate_LH}
		S_{\rm av} &= k_{\rm av} n_{\rm RE} \frac{e^2}{m_{\rm e} \ln \Lambda_{\rm c}} \frac{n_{\rm e}^{\rm tot}}{n_{\rm e}} \frac{E_\parallel - E_{\rm c}^{\rm eff}}{\sqrt{4 + \bar\nu_{\rm s}(p_\star) \bar\nu_{\rm D}(p_\star)}} ~.
	\end{align}
See reference~\cite{Hesslow18} for a description of the effective critical field $E_{\rm c}^{\rm eff}$, the slowing-down frequency $\bar\nu_{\rm s}$ and the generalized deflection frequency $\bar\nu_{\rm D}$. Since the effective critical momentum $p_\star$ for runaway generation is itself a function of $\bar\nu_{\rm s}$ and $\bar\nu_{\rm D}$ through $p_\star = \sqrt[4]{\bar\nu_{\rm s}(p_\star)\bar\nu_{\rm D}(p_\star)}/\sqrt{E_\parallel/E_{\rm c}}$, a closed form of equation~\eqref{eq:Avalanche_growth_rate_LH} cannot be given.

In contrast to the increase of the avalanche growth rate, the steady-state flux of electrons into the runaway region as obtained by kinetic simulations with the linearized Fokker-Planck solver CODE \cite{Stahl16} is noticeably reduced in the presence of partially ionized impurity ions for $E/E_{\rm D} \lesssim 0.1$ \cite{Hesslow18_2}. Albeit useful, application of kinetic solvers in 1D transport simulations for the calculation of RE growth rates is unfeasible due to computational costs. Therefore a neural network for the calculation of the Dreicer growth rate has been created recently from CODE simulations covering an experimentally relevant region of parameter space  using a set of 8 input parameters \cite{Hesslow19_2}. Although training of the network was carried out using only Ar and Ne impurities, input parameters were chosen sufficiently general to allow for application to other impurity species. The neural network obtained shows good agreement with CODE results and may therefore be applied in fluid simulations \cite{Hesslow19_2}.

\hypertarget{model:REGIA}{For} the calculation of Dreicer and avalanche growth rates inside \texttt{ASTRA}, the models discussed are incorporated into a stand-alone \texttt{Fortran} module\footnote{The \texttt{Fortran} module for the calculation of runaway electron Dreicer and avalanche growth rates is available at \href{https://github.com/o-linder/runawayelectrongeneration}{https://github.com/o-linder/runawayelectrongeneration}.} and can therefore be used in any \texttt{Fortran} program. A wrapper routine, referred to as \texttt{REGIA} (\texttt{R}unaway \texttt{E}lectron \texttt{G}eneration \texttt{In} \texttt{Astra}), is employed in \texttt{ASTRA} to call the requested module routines and convert quantities to \texttt{ASTRA} units. Since the expressions by Hesslow et al \cite{Hesslow19,Hesslow19_2} provide an improved description of Dreicer and avalanche generation over the classical formulae, both the neural network for $S_{\rm D}$ and equation~\eqref{eq:Avalanche_growth_rate_LH} for $S_{\rm av}$ are applied throughout this study. The implementation of both models was verified directly against the work of Hesslow et at \cite{Hesslow19,Hesslow19_2}. For the calculation of impurity related quantities, \texttt{REGIA} selects from the impurity densities obtained through \texttt{STRAHL} all charge states $l$ which contribute noticeably to the overall electron density or are present in noticeable amounts, i.e. $n_l \cdot \max\left( Z_l, 1\right) \geq 10^{-4} ~ n_{\rm e}$. 

The possibility to use the analytical expressions of Connor \& Hastie \cite{Connor75} and Rosenbluth \& Putvinski \cite{Rosenbluth97}, i.e. equations~\eqref{eq:Dreicer_growth_rate_CH} and \eqref{eq:Avalanche_growth_rate_RP}, is kept to allow for an assessment of the differences between these models and hence of the importance of partially ionized impurities on the RE population obtained in self-consistent MGI simulations. The implementation of the analytical Dreicer and avalanche sources from equations~\eqref{eq:Dreicer_growth_rate_CH} and \eqref{eq:Avalanche_growth_rate_RP} was verified successfully against calculations by the disruption code \texttt{GO} \cite{Papp13}. For this purpose, simulations of a thermal collapse with prescribed exponential temperature decay on time scales $\tau$ to post-disruption temperatures $T_\mathrm{e,f}$ in a clean plasma were carried out. The RE currents obtained by both tools show reasonable agreement within a few \% in the region of parameter space $(\tau, T_{\rm e,f})$ relevant for disruptions in AUG. 
 
The transport of runaways across magnetic flux surfaces is set to occur diffusive with $D = 10^{-3}~{\rm m}^2~{\rm s}^{-1}$ to ensure numerical stability of the simulations without impacting simulation results. Rapid radial transport due to stochastic magnetic field lines during magnetic reconnection is not considered, as the bulk of the RE population is generated only once nested flux surfaces have re-emerged. Other transport mechanisms are expected to be of minor importance, given the fast time scale of the current quench (CQ) and hence RE generation within a few ms in AUG. RE dissipation mechanisms are thus considered important only after the CQ during the RE beam phase of a disruption. However, this aspect is not covered in this work.

From the RE density $n_{\rm RE}$ obtained by \texttt{ASTRA} through evaluation of the transport equation (equation~\eqref{eq:astra_transport_equation}) under consideration of sources $S_j$ and transport $D$, the runaway carried current density is constructed, $j_{\rm RE} = c\hspace*{1pt} e \hspace*{1pt} n_{\rm RE}$, and added to Ohmic current density and bootstrap current density to construct the total plasma current density, i.e. $j_\mathrm{p} = j_\Omega + j_\mathrm{BS} + j_\mathrm{RE}$, thus affecting $\Psi$-evolution in \texttt{ASTRA}. Hereby, the REs are assumed to travel with the speed of light, which gives a less than 1\% deviation for REs with $E_\mathrm{kin} > 6.1 \hspace*{2pt} m_\mathrm{e}\hspace*{1pt}c^2 = 3.1 \hspace*{1pt} \mathrm{MeV}$. This assumption may not hold in the early phase of a disruption as electrons have to perform in excess of $10^{3}$ revolutions in the tokamak 
to extract these amounts of energy from the strong induced electric fields. Simultaneously, the RE contribution to the total current and hence impact on $\Psi$-evolution is negligible. Only at a later stage of the CQ, sufficient amounts of REs are present to noticeably impact $\Psi$-evolution. At this stage, however, REs are expected to have gained enough energy from the electric field to justify the assumption that REs travel with the speed of light.

\section{Experimental scenario}
\label{sec:experimental_scenario}
Simulations of the interplay between deuterium plasma, impurity MGI and REs are performed for AUG discharge \#33108. This particular discharge was tailored for the study of RE generation and mitigation and has been used as a reference case for proceeding RE experiments in multiple experimental campaigns.

In the discharge chosen, Ar is injected into an L-mode limiter plasma with an initial plasma current of 763~kA 
and an applied on-axis magnetic field of 2.5~T. Vertical stability of the plasma during the artificial disruption is facilitated through the circular shape of the plasma. The pre-disruption plasma density was kept low with an average of $\left\langle n_{\rm e} \right\rangle = 2.84 \times 10^{19}~\mathrm{m}^{-3}$ 
to reduce the collisional drag on (seed) runaways and thus to facilitate generation of a substantial RE beam during the disruption. Current conversion was further enhanced by applying 2.625~MW of on-axis ECRH provided by four of the gyrotrons in the last 100~ms prior to MGI, i.e. between $t=0.9~{\rm s}$ and $t=1.0~{\rm s}$.\footnote{In this section, the time $t$ is given with respect to the beginning of the discharge.} 
In the process, a strongly peaked electron temperature profile is established, with on-axis temperatures in excess of 10~keV. As a result, the Dreicer electric field $E_{\rm D}$ is reduced in the central plasma, thus enhancing the primary source of runaways through the exponential sensitivity on $-E_{\rm D}/E_\parallel$ (see equation~\eqref{eq:Dreicer_growth_rate_CH}) and contributing to generation of a considerable RE beam during the disruption. The profiles of electron temperature $T_{\rm e}$, electron density $n_{\rm e}$ and safety factor $q$ averaged between $t = 0.95~{\rm s}$ and $t= 1.00~{\rm s}$ are shown in figure~\ref{fig:kinetic_profiles_pre_MGI}. These profiles are set as initial conditions in the simulations discussed in the following section~\ref{sec:simulations}.

\subsection{Diagnostics}

\ifshowfigures %
  \tikzsetnextfilename{01_pre-MGI_profiles}%
  \begin{figure}[tb!]
	\centering
	\includegraphics[scale=1]{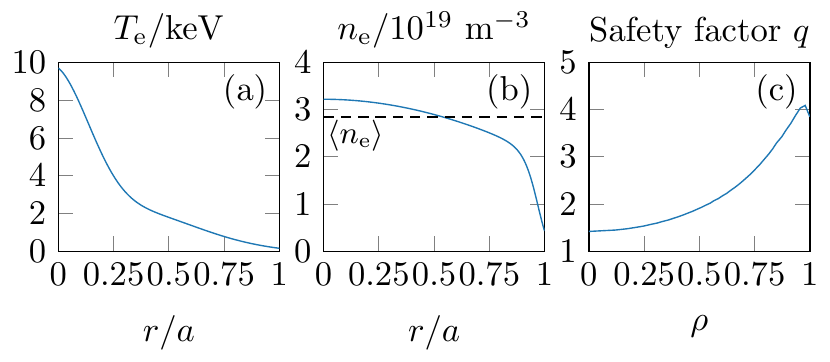}
	\vspace*{-.5cm}
	\caption[Averaged plasma profiles prior to MGI]{\label{fig:kinetic_profiles_pre_MGI} Profiles averaged between $t = 0.95~\rm{s}$ and $t = 1.00~\rm{s}$ prior to Ar MGI in AUG \#33108 of (a) electron temperature $T_{\rm e}$ obtained from electron cyclotron emission (ECE) and Thomson scattering (TS) measurements, (b) electron density $n_{\rm e}$ from TS and interferometry, (c) the safety factor $q$.}
\end{figure}%
 \fi

\ifshowfigures %
  \tikzsetnextfilename{02_diag_geom}%
  \pgfplotsset{select coords between index/.style 2 args={
    x filter/.code={
        \ifnum\coordindex<#1\def\pgfmathresult{}\fi
        \ifnum\coordindex>#2\def\pgfmathresult{}\fi
    }
}}

\begin{figure}
	\centering
	\includegraphics[scale=1]{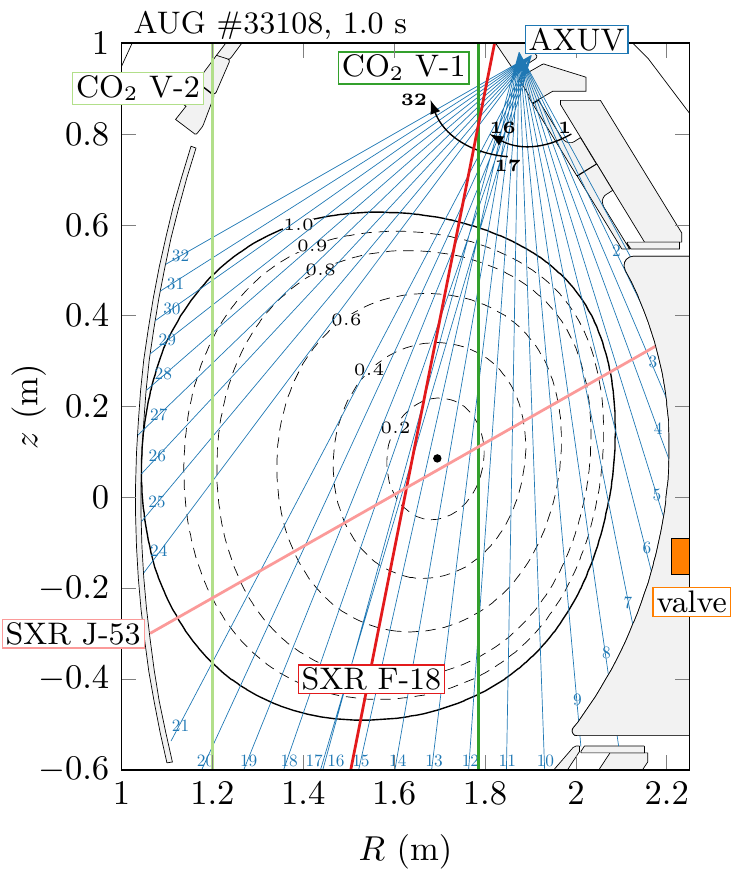}
	\centering
	\caption[AUG \#33108 diagnostics and magnetic equilibrium]{%
		\label{fig:diag_geom}
		Line of sights of the relevant diagnostics for the analysis of AUG \#33108, being two vertical CO$_2$ interferometers (green), the AXUV photodiodes (blue) and two central SXR channels (red). Ar is injected from the LFS valve (orange) in sector 13. The magnetic equilibrium at the onset of Ar MGI at $t_{\rm MGI} = 1.0~\mathrm{s}$ is illustrated by flux surfaces of normalized poloidal flux $\rho_\theta$.}
\end{figure}%
 \fi

The evolution of the plasma following injection of Ar is captured by several diagnostics. The line integrated electron density $\bar{n}_{\rm e}$ is obtained from CO$_{\rm 2}$ interferometers along two vertical lines of sight, covering the central plasma (chord V-1) and the high-field-side (HFS) outer plasma (chord V-2) in sector 11 of the AUG vessel (see figure~\ref{fig:diag_geom}). Radiation measurements are carried out with the Absolute eXtreme UltraViolet diagnostic (AXUV), a diode array consisting of 48 vertical channels (although only the relevant first 32 channels are shown in figure~\ref{fig:diag_geom}), and can be used to reconstruct the propagation of the deposited material. The occurrence of the TQ can be determined by measurements of Soft X-Ray (SXR) radiation in the plasma center using one vertical and one horizontal channel of the available SXR photo diodes.

\subsection{Ar MGI}
In AUG \#33108, Ar is injected from the low-field-side (LFS) valve located in sector 13 (see figure~\ref{fig:diag_geom}) into the vessel. The valve's gas reservoir of 85~cm$^3$ and feed line of 15~cm$^3$ were filled with 0.73~bar of Ar at ambient temperature, corresponding to $1.75\times 10^{21}$ Ar atoms - around 7 times the deuterium content of the confined plasma. 
Injection of material is triggered at $t_{\rm MGI} = 1.0~{\rm s}$. At the same time, ECRH heating is shut off. The flow of Ar out of the valve is reconstructed in 0D under the assumption of ideal flow from the continuity equation $\mathrm{d}N_{\rm Ar}/\mathrm{d}t + v_{\rm Ar} \hs N_{\rm Ar} \hs A_{\rm v}(t)/\hs V_{\rm v} = 0$ (see appendix A of reference~\cite{Pautasso07} for details). A finite opening time of 1~ms of the valve's aperture of size $A_{\rm v}$ is assumed \cite{Fable16}. The Ar flow obtained is illustrated in figure~\ref{fig:experimental_evolution}(a). A more complete treatment of material injection is e.g. achieved by the 1D code IMAGINE \cite{Fil15}. Throughout the remainder of this work, the temporal evolution $t$ will be given with respect to $t_{\rm MGI}$ instead of the beginning of the discharge. 

\ifshowfigures %
  \tikzsetnextfilename{03_evolution}%
  \pgfplotsset{legend image code/.code={
	\draw[mark repeat=2,mark phase=2] plot coordinates {
		(0cm,0cm)
		(0.15cm,0cm)        
		(0.3cm,0cm)         
		};%
	}}

\begin{figure}
	\centering
	\includegraphics[scale=1]{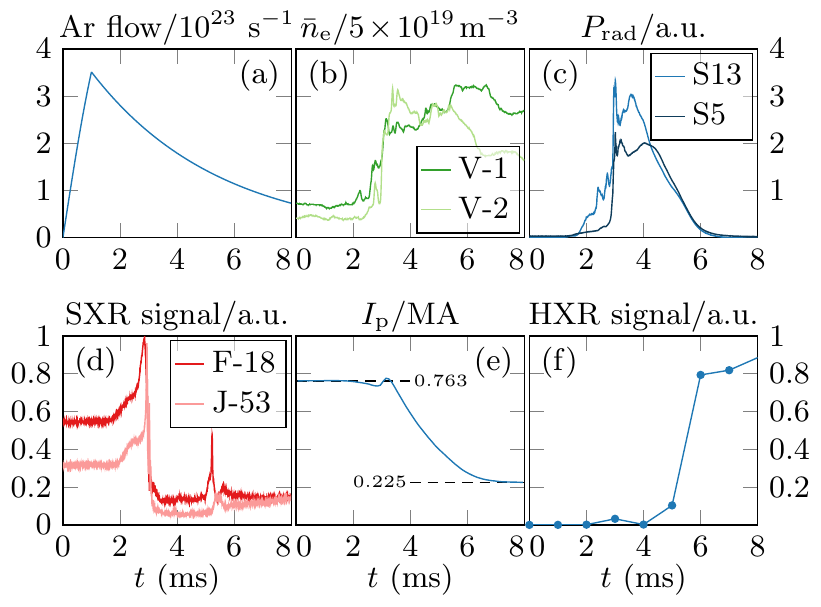}
	\centering
		
	\caption[Temporal evolution of experimentally obtained and derived quantities of AUG \#33108]{%
		\label{fig:experimental_evolution}
		Temporal evolution of experimentally obtained and derived quantities of AUG \#33108 following the MGI trigger at $t_{\rm MGI} = 1.0~{\rm s}$, being (a) the calculated Ar flow from the valve, (b) measurements of the line integrated electron density $\bar{n}_{\rm e}$ by the CO$_2$ interferometers, (c) the total radiated power $P_{\rm rad}$ calculated from AXUV measurements in sectors 13 and 5, (d) the signals of two central SXR measurements, (e) the reconstructed plasma current $I_{\rm p}$ and (f) HXR measurements as an indication for the RE population.
		}
\end{figure}%
 \fi

\ifshowfigures %
  \tikzsetnextfilename{04_bolo_s13_norm}%
  \begin{figure}
	\centering
	\includegraphics[scale=1]{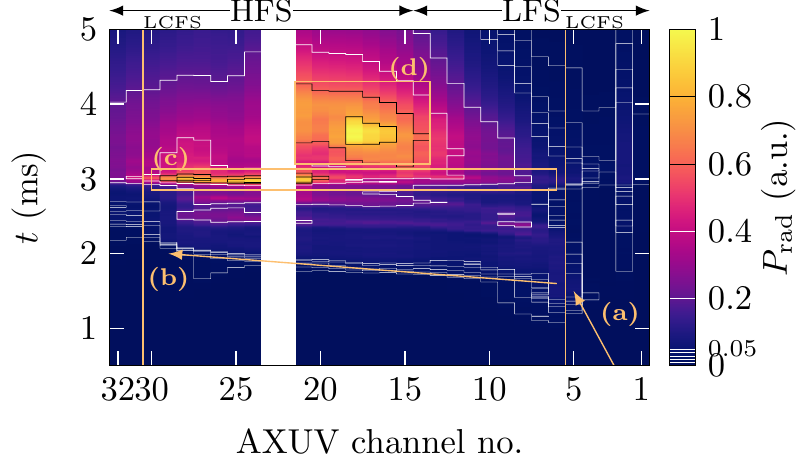}
	\centering
	\caption[Radiation measurements by AXUV diodes in sector 13]{%
		\label{fig:bolo_s13}
		Radiation measurements of the individual channels of the AXUV photodiodes in sector 13 (see figure~\ref{fig:diag_geom} for their line-of-sights). Channels 1 to 14 capture LFS radiation, i.e. their lines-of-sight intersect the equatorial plane beyond the magnetic axis, whereas channels 15 to 32 cover the HFS. Core radiation is captured by channels 6 to 30. Note, that this classification is valid only in the very early phase of the disruption. The Ar injected reaches the LFS-LCFS at $t = 1~{\rm ms}$ (a), ionizes and re-distributes poloidally (b), eventually causing a thermal collapse at $t = 3~{\rm ms}$ (c), followed by strong central radiation (d).}
\end{figure}%
 \fi

\ifshowfigures %
  \tikzsetnextfilename{05_neoclassical_only}%
  \begin{figure*}
	\centering
	\includegraphics[scale=1]{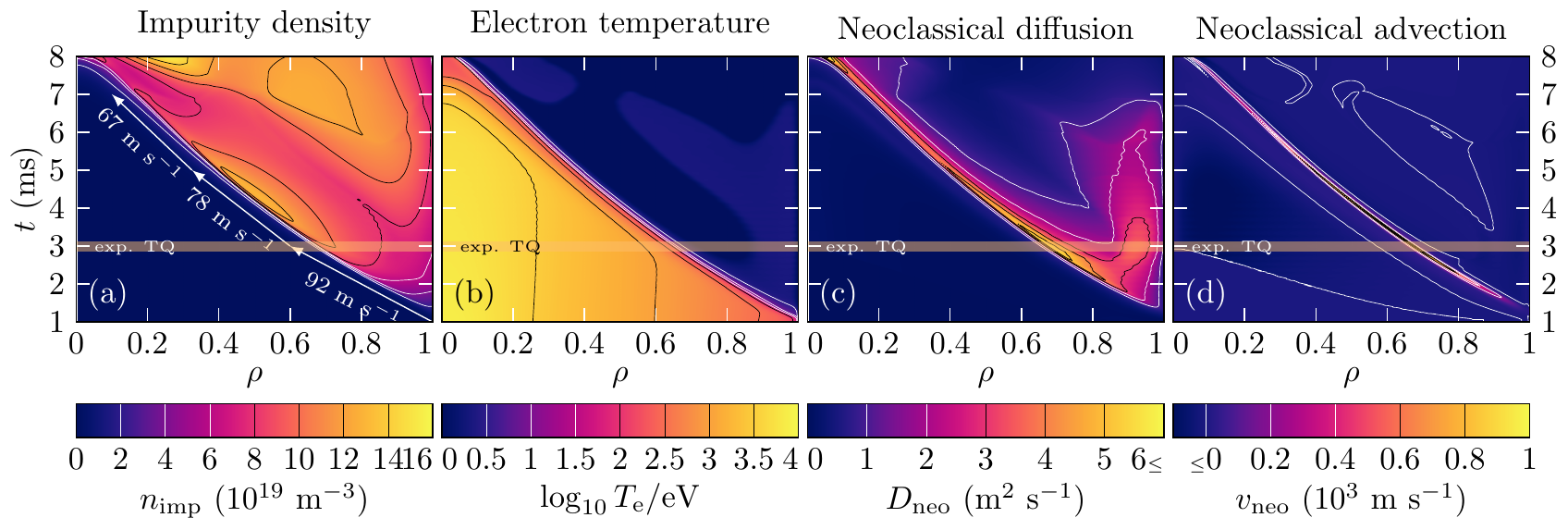}
	\centering
	\caption[Evolution of plasma profiles in Ar MGI simulations with neoclassical transport only]{%
		\label{fig:impurity_evolution_wo_add_transport}%
		Spatio-temporal plasma evolution of AUG \#33108 in a simulation of Ar MGI, where the transport of impurity ions is treated solely neoclassically: (a) impurity density of both neutral and ionized Ar, as well as effective inward propagation velocity $v_{\rm eff}$, (b) logarithmic electron temperature, (c) neoclassical diffusion coefficient and (d) neoclassical drift velocity. 
		In the experiment, the TQ is observed to occur between $t=2.85~\mathrm{ms}$ and $t=3.13~\mathrm{ms}$ (orange ribbon); cf. figure~\ref{fig:experimental_evolution}(d) and figure~\ref{fig:bolo_s13}.
		}
\end{figure*}%
 \fi

The material injected reaches the last closed flux surface (LCFS) at the LFS at $t = 1~{\rm ms}$, as indicated by the increase of UV radiation measured by channels \#5 and \#6 of the AXUV diagnostic in sector 13 (see figure~\ref{fig:bolo_s13}). Over the following 1~ms, the AXUV signal increases across the remaining channels, as part of the Ar gas ionizes and redistributes poloidally. Ar propagates further inwards (see increase of the line averaged electron density $\bar{n}_{\rm e}$ in figure~\ref{fig:experimental_evolution}(b)) and causes a thermal collapse of the plasma between $t = 2.85~{\rm ms}$ and $t=3.13~{\rm ms}$. The accompanying increase of radiation is detected by the AXUV arrays both in sector 13 and 5 (see figure~\ref{fig:experimental_evolution}(c) and figure~\ref{fig:bolo_s13}), as well as by SXR measurements of the central plasma (see figure~\ref{fig:experimental_evolution}(d)). During the TQ, the plasma current is observed to increase by 41~kA (see figure~\ref{fig:experimental_evolution}(e)), presumably due to magnetic reconnection, triggered by the cold gas front reaching the $q=2$ rational surface \cite{Fable16}. Over the 2.5~ms following the TQ, the remnant core plasma continues to radiate strongly. Simultaneously, the plasma current decreases rapidly from an initial value of 763~kA down to 225~kA. The remaining current is carried by relativistic electrons, as indicated by a strong increase of the signal of the Hard X-Ray (HXR) diagnostic (see figure~\ref{fig:experimental_evolution}(f)).
\section{MGI simulations}
\label{sec:simulations}

\subsection{Scenario specific simulation settings}
Argon MGI in AUG discharge \#33108 is simulated with the model presented in section~\ref{sec:model_description}. The source of Ar is set 1~cm outside the LCFS since the propagation of material from the valve to the core plasma is not modeled. Instead, the experimentally observed delay of 1~ms between valve trigger and detection of Ar at the LFS-LCFS (see figure~\ref{fig:bolo_s13}) is applied to the calculated flow rate of Ar from the valve (see figure~\ref{fig:experimental_evolution}(a)) and used as source rate in the simulation. The neutral Ar injected propagates into the core plasma with thermal velocity, being $v_{\rm th} = \sqrt{T/m} = 246~\mathrm{m}~\mathrm{s}^{-1}$.

The simulations presented in the following are carried out using a numerical grid consisting of 201 points inside \texttt{ASTRA}. A minimum and maximum time step of $10^{-5}~\mathrm{ms}$ and $1~\mathrm{ms}$ respectively are chosen to capture the transient dynamics of Ar MGI, while allowing efficient calculation of the later stages of the induced disruption. Applying larger values for the minimum time step impacts simulation results due to exponentiation of the RE population. The time step is varied by \texttt{ASTRA} adaptively throughout the simulation to ensure a maximum change  of the profiles evolved of 1~\% during a single time step. The electron and ion heat transport equations are solved applying a fixed boundary condition of $T_{\rm e} = T_{\rm i} = 0.5~\mathrm{eV}$ to ensure sufficiently large temperatures necessary for equilibrium calculations and thus for the stability of the simulations.

Calculations of the impurity evolution inside \texttt{STRAHL} are performed on the same numerical grid as used in \texttt{ASTRA}, while being expanded outside the LCFS to resolve the Ar source. A constant time step of $10^{-5}~\mathrm{ms}$ is set to ensure resolving the fast dynamics of the atomic processes involved.

\subsection{Impurity Ar evolution}
\label{sec:impurity_ar_evolution}
In the simulations performed, neutral Ar deposited outside the LCFS propagates into the core plasma with thermal velocity, ionizes and radiatively cools down the local plasma to a few eV in the process (see figure~\ref{fig:impurity_evolution_wo_add_transport}(a,b)). As a result of the combined effect of depletion of neutrals in the gas plume through ionization and replenishment through the external source, the effective inward propagation velocity $v_{\rm eff}$ of the neutral gas front is noticeably reduced from the thermal velocity down to on average 78~m~s$^{-1}$. 

The transport of ionized Ar is governed by neoclassical mechanisms in these simulations. At the periphery of the cold gas front, the neoclassical diffusion coefficient increases up to on average $D_{\rm neo} = 4.0~\mathrm{m}^2~\mathrm{s}^{-1}$ (see figure~\ref{fig:impurity_evolution_wo_add_transport}(c)), gradually decreasing towards the edge. Simultaneously, an outward drift of on average $v_{\rm neo} = 540~\mathrm{m}~\mathrm{s}^{-1}$\footnote{Note, that the neoclassical transport coefficients $D_{\rm neo}$ and $v_{\rm neo}$ are the averages of individual transport coefficients from all charge states $i$ with finite density, $n_{\mathrm{imp},i} \geq 10^{3}~\mathrm{m}^{-3}$. Averaged coefficients are shown only for illustrative purposes; transport calculations for individual charge states are performed applying the corresponding coefficients obtained by \texttt{NEOART}.}%
occurs in the region of largest impurity density gradients, thus being highly localized (see figure~\ref{fig:impurity_evolution_wo_add_transport}(d)). The resulting neoclassical fluxes are in consequence noticeably smaller than the influx of neutrals. The inward propagation of the cooling front is therefore primarily driven by incoming neutrals.

Under consideration of aforementioned impurity transport mechanisms, the plasma stored energy is lost on longer time scales then experimentally observed. Whereas the TQ is detected to occur between $t=2.85~\mathrm{ms}$ and $t=3.13~\mathrm{ms}$ in AUG \#33108 (see figure~\ref{fig:experimental_evolution}(d)), the plasma stored energy is gradually lost over the duration of 5.2~ms in the simulation performed, with the central electron temperature collapsing only at around $t = 6.3~\mathrm{ms}$ (see figure~\ref{fig:impurity_evolution_wo_add_transport}(b)). Similarly, the sudden rise of the line integrated electron density prior to the TQ (see figure~\ref{fig:experimental_evolution}(b)) cannot be captured. Hence, the penetration of material into the plasma core occurs too slow under consideration of only neutral propagation with thermal velocity and neoclassical impurity transport to reproduce experimental observations. Consequently, additional transport mechanisms are considered to dominate impurity and plasma evolution.

To demonstrate the necessity for the consideration of additional transport mechanisms, the inward propagation of neutral Ar is unrealistically assumed to occur with $v_{\rm th} = 1000~\mathrm{m}~\mathrm{s}^{-1}$ in a following simulation; corresponding to a gas temperature of 4800~K in the valve. Yet under these conditions, ionization of incoming neutrals reduces the average effective propagation velocity of the gas plume down to on average $v_{\rm eff} = 114~\mathrm{m}~\mathrm{s}^{-1}$. Being not significantly faster than the effective propagation velocity calculated for the simulation of Ar at room temperature, the plasma evolution obtained is qualitatively similar for both values of the valve temperature. The choice of the initial neutral velocity has therefore no significant influence on the occurrence of a fast TQ (or rather lack of therefore) in the simulations, advocating the consideration of additional transport mechanisms.

\subsection{Impurity mixing}
\label{sec:impurity_mixing}
The experimentally observed plasma response to Ar MGI can be reproduced by impurity transport simulations under the assumption of rapid redistribution of ionized material (see figures~\ref{fig:impurity_evolution_mixing_profiles} and \ref{fig:impurity_evolution_mixing_traces}). In AUG \#33108, an $(m,n) = (2,1)$ mode and higher harmonics are excited as the cold gas front reaches the $q=2$ surface \cite{Fable16}, eventually triggering magnetic reconnection and resulting in rapid redistribution of impurities inside the $q=2$ surface.

\ifshowfigures %
  \tikzsetnextfilename{06_impurity_mixing_profiles}%
  \begin{figure}[t!]
	\centering
	\includegraphics[scale=1]{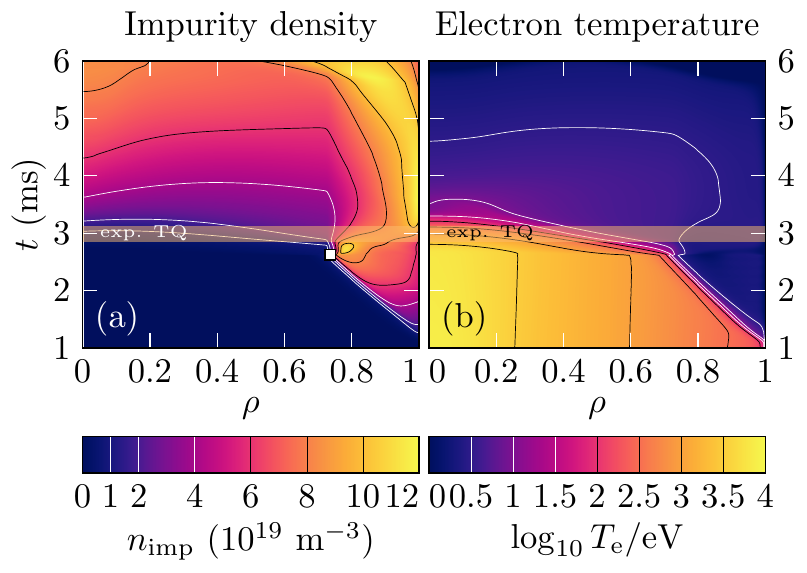}
	\centering
	\vspace*{-.2cm}
	\caption[Evolution of plasma profiles in Ar MGI simulations with impurity mixing]{%
		\label{fig:impurity_evolution_mixing_profiles}%
		Spatio-temporal plasma evolution of AUG \#33108 in a simulation of Ar MGI with impurity mixing: (a) impurity density of both neutral and ionized Ar, (b) logarithmic electron temperature. The neutral Ar deposited penetrates the core plasma up to $\rho = 0.74$ at $t = 2.6~\mathrm{ms}$ (white square in (a)).
		}
\end{figure}%
 \fi 

\ifshowfigures %
  \tikzsetnextfilename{07_impurity_mixing_traces}%
  \pgfplotsset{legend image code/.code={
	\draw[mark repeat=2,mark phase=2] plot coordinates {
		(0cm,0cm)
		(0.15cm,0cm)        
		(0.3cm,0cm)         
		};%
	}}

\begin{figure}[t!]
	\centering
	\includegraphics[scale=1]{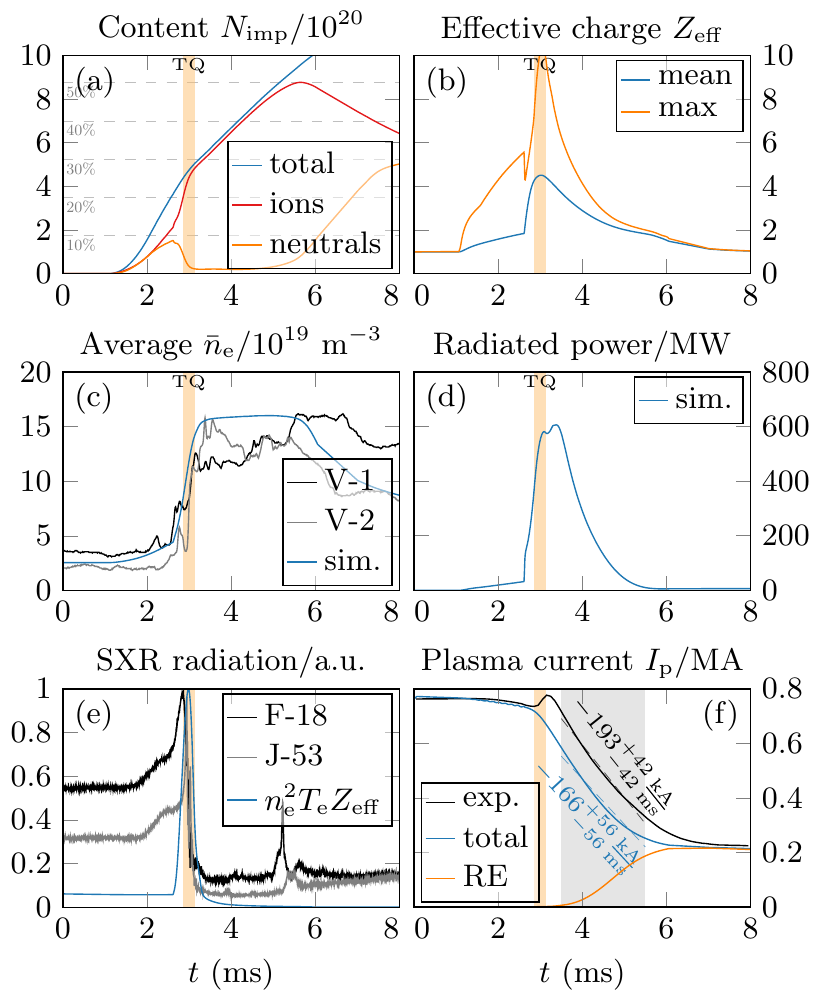}
	\centering
	\vspace*{-.5cm}
	\caption[Evolution of plasma traces in Ar MGI simulations with impurity mixing]{%
		\label{fig:impurity_evolution_mixing_traces}
			Temporal evolution of plasma parameters in an Ar MGI simulation with impurity mixing: (a) content of the total, ionized and neutral impurity Ar inside the LCFS, (b) volume averaged and maximum effective charge, (c) line averaged electron density compared to measurements of the two CO$_2$ chords, (d) radiated power, (e) line averaged value of $n_{\rm e}^2T_{\rm e}Z_{\rm eff}$ as proxy for the measured SXR radiation \cite{Igochine10} to determine the onset of the TQ\footnotemark compared to two central channels of the SXR diagnostic, (f) total and RE plasma current compared to experimental reconstruction. The rate of current decay is averaged between $t=3.5~\mathrm{ms}$ and $t=5.5~\mathrm{ms}$ (gray region).
		}
\end{figure}%
 \fi %
\footnotetext{Note, that the onset of the TQ cannot be determined from $T_{\rm e}$ measurements. 
The ECE signal is in high-density cutoff; the temporal resolution of TS is insufficient.}

The rapid redistribution of impurity ions due to MHD phenomena can be mimicked in 1D transport simulations by applying increased transport coefficients inside the $q=2$ surface to the impurity ion population when the cold gas front reaches this particular rational surface at $t_{q=2}$. %
%
%
Since the process of magnetic reconnection is a transient phenomenon, the transport coefficients prescribed decrease exponentially on time scales $\tau_{\rm add}$, i.e. in the case of the diffusion coefficient $D_{\rm add}(t) = D_{\rm add}^{\rm max} \exp \left( - \{t - t_{q=2}\}/\tau_{\rm add} \right) \cdot \Theta(t-t_{q=2})$ and similarly for the pinch velocity. In the simulations presented, $D_{\rm add}^{\rm max} = 100~\mathrm{m}^2~\mathrm{s}^{-1}$,  $v_{\rm add}^{\rm max} = -200~\mathrm{m}~\mathrm{s}^{-1}$ and $\tau_{\rm add} = 1.0~\mathrm{ms}$ are applied. Outside the $q=2$ surface, no additional transport is applied. The numerical values for the additional transport coefficients are obtained by matching simulation results with experimental observations of the line integrated electron density $\bar{n}_{\rm e}$ and plasma current evolution $I_{\rm p}$ (see figures~\ref{fig:experimental_evolution}(b) and (e)). The heat transport coefficient is additionally increased by $\chi_{\rm add} = 10~\mathrm{m}^2~\mathrm{s}^{-1}$ for both electrons and ions in the entire simulation domain to prevent the occurrence of strongly localized hot plasma beamlets in the simulation during this phase of the discharge. Compared to neoclassical impurity transport discussed (see figure~\ref{fig:impurity_evolution_wo_add_transport}(c)), the additional transport is significantly increased.

Alternatively to the rapid redistribution of impurities, the TQ may be triggered by assuming significantly increased heat transport ($\chi > 100~\mathrm{m}^2~\mathrm{s}^{-1}$) inside the $q=2$ surface. Even though the loss of thermal energy occurs significantly faster in simulations of this case, both the obtained line integrated electron density $\bar{n}_{\rm e}$ and the plasma current decay rate $\left|\mathrm{d}I_{\rm p}/\mathrm{d}t\right|$ fall noticeably below experimental measurements. Consequently, the rapid redistribution of impurity ions is a necessary assumption to reproduce experimental observations of AUG \#33108.

On a side note, another manifestation of the MHD phenomena triggered by the cold gas front is the occurrence of a sudden increase of the total plasma current (see figure~\ref{fig:experimental_evolution}(e)). However being focused on the simulation of material propagation and RE generation, reproducing the increase in total plasma current in the simulations goes beyond the scope of this work. Consequently, the evolution of the total plasma current cannot be captured by the simulations performed. However, the current decay rate $\mathrm{d}I_{\rm p}/\mathrm{d}t$ is well reproduced during the CQ (see figure~\ref{fig:impurity_evolution_mixing_traces}(f)).

\subsection{Plasma evolution with rapid Ar redistribution}
In the presence of additional transport coefficients, experimental observations are reproduced by the Ar MGI simulation carried out (see figures~\ref{fig:impurity_evolution_mixing_profiles} and \ref{fig:impurity_evolution_mixing_traces}). The neutral Ar deposited outside the LCFS propagates into the core plasma. Upon reaching the $q=2$ surface at $t_{q=2} = 2.62~\mathrm{ms}$, the additional impurity transport coefficients prescribed to mimic the impact of magnetic reconnection rapidly transport ionized Ar into the central plasma. A considerable amount of impurity ions, $n_{\rm Ar} \geq 10^{19}~\mathrm{m}^{-3}$, is present at the magnetic axis after 0.42~ms at $t \geq 3.04~\mathrm{ms}$.

Given the exponential decay of the additional transport coefficients on a time scale of $\tau_{\rm add} = 1~\mathrm{ms}$, the associated transport alone is insufficient to obtain the level of central impurity accumulation necessary to reproduce experimental observations of the artificial disruption. Here, neoclassical impurity transport provides a substantial contribution to redistribute impurities in the plasma center. Omitting this mechanism, the impurity content inside the $q=2$ surface is reduced by as much as 60\% during rapid impurity redistribution. Even though sufficient to induce a fast TQ on experimental time scales and to reproduce the increase of the line averaged electron density observed, the plasma current decays too slow under these conditions. This suggests, that in the experiment, a significant fraction of the impurity ions are present in the central plasma, where the remnant plasma current is located, thus accelerating current decay. Still, in the absence of neoclassical transport processes, experimental conditions may be recovered by increasing the additional transport coefficients applied, e.g. $D_{\rm add}^{\rm max} = 100~\mathrm{m}^2~\mathrm{s}^{-1} \to 200~\mathrm{m}^2~\mathrm{s}^{-1}$ and $v_{\rm add}^{\rm max} = -200~\mathrm{m}~\mathrm{s}^{-1} \to -1000~\mathrm{m}~\mathrm{s}^{-1}$, thus emphasizing the importance of neoclassical impurity transport for impurity redistribution inside the $q=2$ surface.

Even though neoclassical processes alone are also insufficient to cause a thermal collapse of the plasma on experimentally observed time scales (see section~\ref{sec:impurity_ar_evolution}), the rapid redistribution of impurity ions through application of additional transport coefficients modifies the plasma profiles favorably for inward neoclassical transport. As the steepness of the ion density gradient greatly decreases in the process of redistribution, the magnitude of the outward neoclassical pinch is significantly reduced, even reversing direction (with low magnitude). Furthermore, neoclassical diffusion contributions are considerably increased compared to the case without additional transport (see figure~\ref{fig:impurity_evolution_wo_add_transport}(c)), being on average inside the $q=2$ surface of the order of $\left\langle D_{\rm neo} \right\rangle = 20~\mathrm{m}^2~\mathrm{s}^{-1}$. 

In the process of propagating into the hot central plasma, the Ar is ionized to high ionization states, reaching in the vicinity of the magnetic axis up to $Z_{\rm imp} = 12.8$ and rapidly increasing the density of plasma electrons (see figures~\ref{fig:impurity_evolution_mixing_traces}(b) and (c)). Given the time scale of material deposition and penetration, numerical evolution of the individual impurity charge state densities from rate equations is mandatory as the charge state distribution is not in steady state (coronal equilibrium) throughout the simulation (see figure~\ref{fig:impurity_fraction}). Evaluating the steady state impurity ion distribution for the electron temperature evolution obtained in self-consistent simulations, the Ar population adapts too quickly to changing plasma conditions as compared to simulation results. In regions of hot plasma, i.e. in front of the cold gas front and inside the $q=2$ surface during rapid redistribution, impurity ionization is significantly overestimated. Particularly overpopulated are the ionization stages Ar$^{8+}$ (Ne-like Ar) and Ar$^{16+}$ (He-like Ar). Similarly, recombination to singly ionized Ar in the wake of the cold gas front and in the post-TQ plasma occurs too fast. Consequently, application of the steady state Ar charge state distribution is not suitable to describe the charge state evolution in MGI simulations.

As a consequence of the rapid redistribution of impurities, the majority of the plasma thermal energy is lost through strong radiation within $0.68~\mathrm{ms}$ between $t = t_{q=2}$ and $t = 3.30~\mathrm{ms}$. In the resulting cold post-TQ plasma of a few eV, the Ar recombines to low ionization stages, thus decreasing the free electron density in the central plasma. Yet, neutral Ar deposited in the core plasma following the TQ still becomes ionized in the outer half of the plasma, thus keeping the line averaged electron density approximately constant in the 2.43~ms following the TQ. The plasma temperature of a few eV is maintained throughout this phase despite further radiative losses through Ohmic heating by the remaining, decaying plasma current as a result of strong parallel electric fields induced. Only after the majority of the Ohmic current has decayed at $t = 5.72~\mathrm{ms}$, the remainder of the plasma stored energy is lost through impurity radiation and the plasma temperature approaches 1~eV. In this environment, Ar recombination back to neutral particles occurs, decreasing the line averaged electron density.

The plasma evolution described agrees qualitatively with experimental observations of AUG \#33108. By assuming rapid redistribution of ionized Ar inside the $q=2$ surface, the magnitude and time scale of the increase in line averaged electron density following Ar MGI as detected by the two CO$_2$ interferometers is reproduced in the simulation (see figure~\ref{fig:impurity_evolution_mixing_traces}(c)). Similarly, the duration of the TQ as reconstructed from the evolution of the line averaged quantity $n_\mathrm{e}^2 T_{\rm e} Z_{\rm eff}$ as a proxy for SXR radiation matches estimates from the SXR diagnostics (see figure~\ref{fig:impurity_evolution_mixing_traces}(e)). Yet, the onset of the TQ obtained in the simulation is slightly delayed by $0.12~\mathrm{ms}$ with respect to experimental estimates. During the CQ following, the averaged current decay rate of $-166_{-56}^{+56}~\mathrm{kA}/\mathrm{ms}$ calculated agrees with experimental reconstructions of $-193_{-42}^{+42}~\mathrm{kA}/\mathrm{ms}$ within errorbars, both being averaged between $t=3.5~\mathrm{ms}$ and $t=5.5~\mathrm{ms}$ (see figure~\ref{fig:impurity_evolution_mixing_traces}(f)). Here, the spike of the total plasma current observed experimentally might impact the rate of current decay. Due to the selection procedure of the additional transport coefficients $D_{\rm add}^{\rm max}$ and $v_{\rm add}^{\rm max}$, the post-CQ RE current measured in AUG \#33108 is matched by the simulation presented. As a result, the plasma evolution during Ar MGI is captured by the simulation throughout the different phases of the disruption.

\ifshowfigures %
  \tikzsetnextfilename{08_impurity_fraction}%
  \begin{figure}
	\centering
	\includegraphics[scale=1]{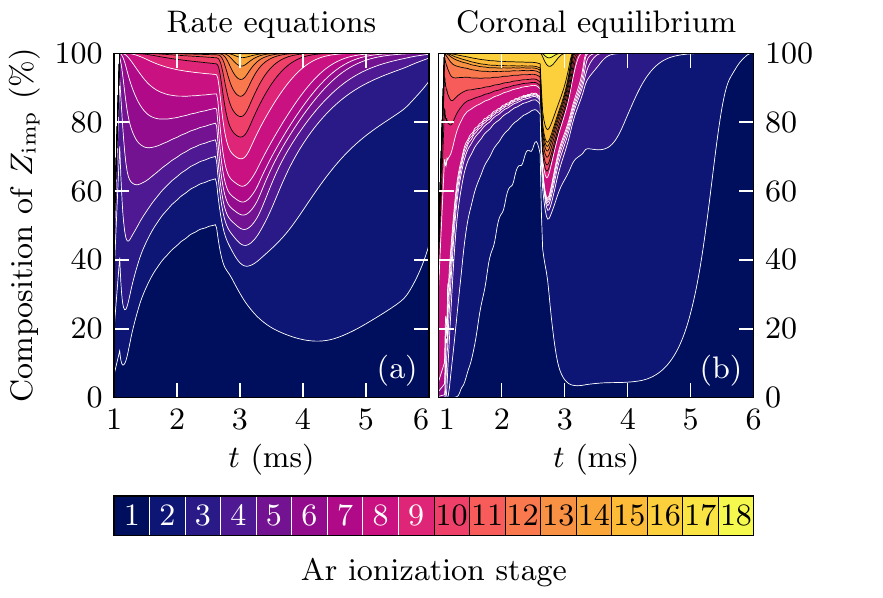}
	\centering
	\caption[Contribution of individual charge states to the average ionization state]{%
		\label{fig:impurity_fraction}%
		Normalized contributions $\int Z_j n_j ~\mathrm{d}V/\sum_i \int Z_i n_i ~ \mathrm{d}V$ of individual Ar charge states $j$ to the global average impurity ion ionization state $Z_{\rm imp} = \sum_i \int Z_i n_i ~\mathrm{d}V/\sum_i \int n_i ~\mathrm{d}V$ in regions where $\sum_i n_i \geq 10^{16}~\mathrm{m}^{-3}$ obtained (a) through evolution of the charge state distribution according to rate equations and (b) through evaluation of the steady state charge state distribution (coronal equilibrium) according to the electron temperature evolution obtained in case (a). Impurity Ar is injected into the vessel at $t=1~\mathrm{ms}$.
		}
\end{figure}%
 \fi

\ifshowfigures %
  \tikzsetnextfilename{09_RE_response}%
  \begin{figure*}
	\centering
	\includegraphics[scale=1]{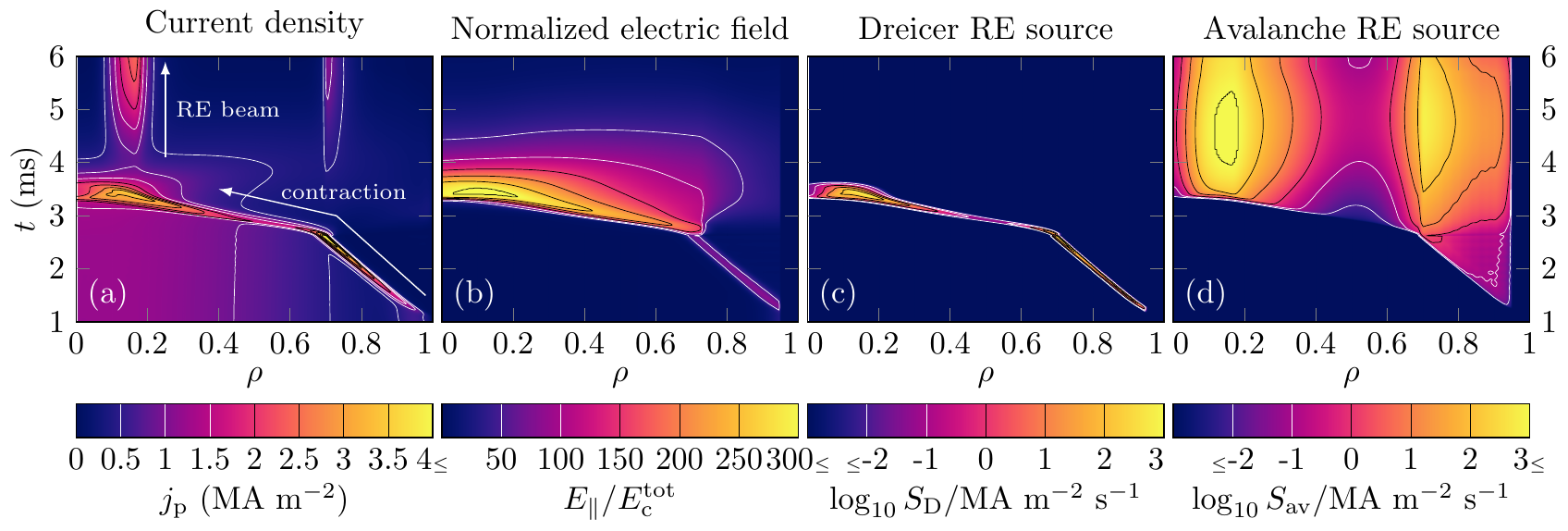}
	\centering
	\caption[RE response to Ar MGI in simulations of AUG \#33108]{%
		\label{fig:RE_response}%
		Spatio-temporal response of RE related quantities to Ar MGI in simulations of AUG \#33108: (a) total plasma current density, (b) parallel electric field normalized to the total critical electric field $E_{\rm c}^{\rm tot} = E_{\rm c}\hspace*{1pt}n_{\rm e}^{\rm tot}/n_{\rm e}$, (c) logarithmic Dreicer RE source and (d) logarithmic avalanche RE source. The parallel electric field normalized to the Dreicer field, $E_\parallel/E_{\rm D}$, evolves similar as the logarithmic Dreicer source and is thus not shown.
		}
\end{figure*}%
 \fi

\subsection{RE evolution}
Throughout the artificially induced disruption by Ar MGI discussed in the previous section, the generation of REs is simulated (see figure~\ref{fig:RE_response}). The source mechanisms considered are highly sensitive to the evolution of the plasma kinetic profiles through (in a simple, analytical picture) exponential dependence on the electron temperature in the case of primary generation, i.e. $S_{\rm D} \propto \exp(-T_{\rm e})$ (see equation~\eqref{eq:Dreicer_growth_rate_CH}), and on the electron density in the case of avalanche multiplication, i.e. $n_{\rm RE}(t) \propto \exp(-n_{\rm e})$ (see equation~\eqref{eq:Avalanche_growth_rate_RP}). Therefore, the deposition and propagation of impurity ions in the plasma core strongly influences the amount of REs generated in the process. Yet, the final RE current can be reproduced in MGI simulations with reasonable agreement to experimental observations if the additional transport coefficients prescribed are chosen such that the evolution of the line averaged electron density is well approximated in the simulation.

Analyzing the response of the plasma and RE current densities to Ar MGI in the simulation presented (see figure~\ref{fig:RE_response}), the edge Ohmic current starts to contract as the material propagates into the confined core; cooling down the plasma to a few eV and reducing the plasma conductivity in the process. Simultaneously, the local electric field $E_\parallel$ increases significantly to counter the change in magnetic flux. Even though increasing well above levels $E_\parallel/E_{\rm D}$ for substantial primary RE generation, these conditions do not last locally sufficiently long to generate relevant amounts of primary RE at a given radial location, let alone secondary REs through avalanching (see $\rho \gtrsim 0.8$ in figure~\ref{fig:RE_response}). Note that the simulations presented do not employ a RE seed population.

Only with prescription of additional impurity transport coefficients as the neutral gas front reaches the $q=2$ surface, noticeable amounts of REs are generated. With increased propagation of the cooling front towards the plasma center, contraction of the current density is accelerated, inducing increasingly stronger electric fields. Given the eventual strong localization of Ohmic current density in the plasma center, the net power loss through impurity radiation is considerably decreased by Ohmic heating in the vicinity of $\rho = 0.16$, keeping the electron temperature in this region transiently between 10~eV and 20~eV for $0.21~\mathrm{ms}$ (see figure~\ref{fig:impurity_evolution_mixing_profiles}(b)). In this environment, sufficient amounts of primary REs are generated to achieve significant avalanche multiplication in the presence of the still strong electric fields, $E_\parallel \gg E_{\rm c}$, of the cold post-TQ plasma of a few eV. This effect is also observed in the vicinity of $\rho = 0.70$. As a result of the selection procedure for the additional transport coefficients, the final RE current is well reproduced by the simulation, being 217~kA, i.e. 4\% below the experimental value $I_\mathrm{RE}^{\rm exp} = 225~\mathrm{kA}$. Given the greater strength of the avalanche source compared to the Dreicer source (see figures~\ref{fig:RE_response}(c) and (d)), the post-CQ RE current consists almost entirely of secondary REs, constituting 99\% of the total RE current. The Dreicer mechanism provides only a seed population for avalanche multiplication of REs. 

The simulation discussed was performed calculating both the primary and secondary RE source from expressions by Hesslow et al \cite{Hesslow19,Hesslow19_2}. Employing the analytical formulae of Connor \& Hastie \cite{Connor75} and Rosenbluth \& Putvinski \cite{Rosenbluth97} instead  (equations~\eqref{eq:Dreicer_growth_rate_CH} and \eqref{eq:Avalanche_growth_rate_RP}), a larger RE current is obtained at the end of the CQ, being 295~kA, i.e. $+31\%~I_{\rm RE}^{\rm exp}$. In this case, the Dreicer mechanism not only provides a seed population for avalanche multiplication, but contributes a substantial fraction of 35\% to the total RE current. Furthermore, a noticeable RE current is obtained earlier in the simulation, exceeding 1~kA at $t = 1.36~\mathrm{ms}$ as opposed to $t = 2.55~\mathrm{ms}$ when employing the \texttt{CODE} neural network instead. The associated considerable increase of the Dreicer growth rate under application of the analytical expression of equation~\eqref{eq:Dreicer_growth_rate_CH} \cite{Connor75} is expected as several effects reducing the flux of thermal electrons into the runaway region of momentum space are not considered in this model, being partial screening of impurities, the energy dependence of the Coulomb logarithm and radiation back reactions \cite{Hesslow19_2}.

Simultaneously, the strength of the avalanche source is reduced applying the expression of Rosenbluth \& Putvinski \cite{Rosenbluth97} of equation~\eqref{eq:Avalanche_growth_rate_RP}, as the nuclear charge of impurity ions is assumed to be screened completely by bound electrons in this model. However as discussed in reference \cite{Hesslow18_2}, collision rates are enhanced in the presence of partially ionized impurities, leading to a stronger than linear dependence of $S_{\rm av}$ on $E_\parallel$ for large electric fields under consideration of both free and bound electrons, thus increasing the growth rate beyond levels predicted by equation~\eqref{eq:Avalanche_growth_rate_RP}. Nevertheless, despite a larger RE seed population and a longer multiplication time in simulations employing the analytical formulae of equations~\eqref{eq:Dreicer_growth_rate_CH} and \eqref{eq:Avalanche_growth_rate_RP}, a comparable avalanche current is obtained in simulations of both cases.

It should be noted, that the additional transport coefficients employed are chosen to match the post-CQ RE current under application of the RE generation models by Hesslow et al\cite{Hesslow19,Hesslow19_2}. However, experimental conditions can also be reproduced describing the RE sources by the analytic expressions of equations~\eqref{eq:Dreicer_growth_rate_CH} and \eqref{eq:Avalanche_growth_rate_RP} by setting e.g. $D_{\rm add}^{\rm max} = 100~\mathrm{m}^2~\mathrm{s}^{-1} \to 200~\mathrm{m}^2~\mathrm{s}^{-1}$ and $v_{\rm add}^{\rm max} = -200~\mathrm{m}~\mathrm{s}^{-1} \to -400~\mathrm{m}~\mathrm{s}^{-1}$. For this choice of parameters, application of the \texttt{CODE} neural network and equation~\eqref{eq:Avalanche_growth_rate_LH} underestimates the RE current generated, being 173~kA ($-23\%~I_\mathrm{RE}^{\rm exp}$). Yet, the overall evolution of the RE current is qualitatively similar for both choices of transport coefficients discussed for each set of the RE generation models. The relative importance of the RE generation mechanisms is unaffected.

Despite employing different models for RE generation and application of varying additional transport coefficients $D_{\rm add}^{\rm max}$ and $v_{\rm add}^{\rm max}$, the final RE currents obtained in the simulations discussed are of similar magnitude, being within $^{+35\%}_{-23\%}I_{\rm RE}^{\rm exp}$, as the available poloidal magnetic flux is dissipated by the total RE generation following the disruption. Consideration of a hot-tail source is hence not expected to modify simulation results significantly. Application of the analytical formulae of Connor \& Hastie \cite{Connor75} and Rosenbluth \& Putvinski \cite{Rosenbluth97} to described RE generation therefore yields a reasonable estimate for the order of magnitude of the post-disruption RE current in MGI scenarios. Still, the evolution of the RE current during the CQ is captured only under application of the models by Hesslow et al \cite{Hesslow19,Hesslow19_2}, where a substantial RE current is obtained after around 4~ms (see figure~\ref{fig:impurity_evolution_mixing_traces}(f)). Experimentally, the hard X-ray signal increases after 5~ms (see figure~\ref{fig:experimental_evolution}(f); note the temporal resolution of 1~ms), being in line with the simulated RE current evolution. Applying the analytical formulae of Connor \& Hastie \cite{Connor75} and Rosenbluth \& Putvinski \cite{Rosenbluth97} instead, a significant RE current is present already at around 3~ms, contrary to experimental observations of HXR radiation.
\section{Conclusion}
\label{sec:conclusions}
In this work, we presented the toolkit \texttt{ASTRA-STRAHL} for self-consistent 1.5D transport simulations of background plasma, impurity species and runaway electrons in MGI scenarios of tokamak plasmas. The model was applied successfully to study the interactions between aforementioned species in simulations of Ar MGI in AUG \#33108, covering all distinct phases of the artificially induced disruption, i.e. the pre-thermal quench, the thermal collapse and the conversion of Ohmic current to runaway electrons. Despite the complexity and the 3D nature of MGI, the evolution of key plasma parameters, such as line integrated electron density and plasma current, can be reproduced in a 1D framework with good agreement compared to experimental observations. The validity of the 1D approach is likely due to the combination of a flux-surface averaged treatment of an arbitrary magnetic geometry by \texttt{ASTRA}, fast equilibration along magnetic field lines and the strong localization of the impurity radiation up to the thermal quench.

The propagation of impurities into the central plasma is driven by rapid redistribution as a result of presumably MHD activity, triggered by the cold gas reaching the $q=2$ rational surface. Still, neoclassical processes provide a non-negligible contribution to inward impurity ion transport. Considering only neoclassical effects, the cold gas front propagates too slow with $v_{\rm eff} \ll v_{\rm th}$ to induce a thermal quench on experimental time scales as propagation is driven primarily by incoming neutrals. The additional transport necessary can be described reasonably well by a simple 0D model of exponentially decaying coefficients for diffusive and convective transport, thus requiring few free parameters. For the simulation of AUG \#33108, the choices $D_{\rm add}^{\rm max} = 100~\mathrm{m}^2~\mathrm{s}^{-1}$, $v_{\rm add}^{\rm max} = -200~\mathrm{m}~\mathrm{s}^{-1}$ and $\tau_{\rm add} = 1~\mathrm{ms}$ were found suitable to reproduce experimental observations. The applicability of these values for simulations of other AUG discharges will have to be investigated in future work.

The generation of REs in AUG \#33108 is described reasonably well in the simulation presented, reproducing the final RE current obtained experimentally. Consideration of the impact of partially ionized impurities on RE generation through application of the models by Hesslow et al \cite{Hesslow19,Hesslow19_2} is necessary to explain both current and HXR measurements. Simulations evolving the RE population based on the commonly used formulae by Connor \& Hastie \cite{Connor75} and Rosenbluth \& Putvinski \cite{Rosenbluth97} cannot capture these experimental observations, highlighting the importance of a thorough treatment of RE generation.

Application of the model presented to Ar MGI in AUG \#33108 demonstrates the suitability of this toolkit for the further study of MMI in tokamak plasmas, in particular of MGI in AUG discharges. Future work will have to investigate if not only individual discharges, but parametric trends observed experimentally can be reproduced. Here, the impact of varying impurity amounts, species and composition on RE generation is to be studied in MGI scenarios. Eventually, the impurity deposition model employed is to be extended by an SPI model, thus allowing MMI simulations of ITER relevant scenarios.

\newpage\ack
\addcontentsline{toc}{section}{Acknowledgments}
This work was supported by the EUROfusion - Theory and Advanced Simulation Coordination (E-TASC). This work has been carried out within the framework of the EUROfusion Consortium and has received funding from the Euratom research and training programme 2014-2018 and 2019-2020 under grant agreement No 633053. The views and opinions expressed herein do not necessarily reflect those of the European Commission.
%
%

\appendix
\section{Finite volume scheme of \texttt{STRAHL}}
\label{sec:appendix_fvm_STRAHL}
The first order central finite difference scheme used in \texttt{STRAHL} is replaced by a vertex centered finite volume discretization of the same order to enable simulations with arbitrary P\'eclet number $\mu$. Starting from the \texttt{STRAHL} transport equation in cylindrical coordinates,
	\begin{align}
		\frac{\partial n}{\partial t} = \frac{1}{r} \frac{\partial}{\partial r} \left( D \frac{\partial n}{\partial r} - v~\hs n \right) + S ~,
	\end{align}
the conservative form of the transport equation for the density $\bar{n}_i(r_i,t) = \int_{\Omega_i} r\hs n(r,t)~\mathrm{d}r/\int_{\Omega_i} r~\mathrm{d}r$ averaged over the cell $\Omega_i = [r_{i-\frac{1}{2}}, r_{i+\frac{1}{2}}]$ is 
	\begin{align}
		\frac{\partial \bar{n}_i}{\partial t} = \frac{2}{r_{i+\frac{1}{2}}^2 - r_{i-\frac{1}{2}}^2} \left[ r \hs D \frac{\mathrm{d}n}{\mathrm{d}r} - r \hs v \hs n \right]_{r_{i-\frac{1}{2}}}^{r_{i+\frac{1}{2}}} + \bar{S}_i ~.
	\end{align}
Quantities at the cell boundaries are evaluated as the average of both cells, i.e. $X_{i\pm\frac{1}{2}} = \left[ X_i + X_{i\pm 1} \right]/2$. For the transport coefficients, the following substitutions are applied:
	\begin{equation}
	\groupequation{
		\begin{aligned}
		\tilde{D}_{i\pm} &= \frac{r_{i\pm\frac{1}{2}}}{g_i\hs \Delta r_{i\pm}} D_{i\pm\frac{1}{2}}~, &
		\tilde{v}_{i\pm} &= \frac{r_{i\pm\frac{1}{2}}}{2\hs g_i} v_{i\pm\frac{1}{2}}, \\
		g_i &= r_{i+\frac{1}{2}}^2-r_{i-\frac{1}{2}}^2~, &
		\Delta r_{i\pm} &= \mp \left( r_i - r_{i\pm 1} \right) ~.
		\end{aligned}
	}
	\end{equation}
The temporal derivative is discretized applying the $\theta$-method, such that $\partial \bar{n}_i/\partial t = \theta\hs F(\bar{n}_i^{j+1}) + ( 1-\theta) \hs F(\bar{n}_i^j)$. Unconditional stability of the scheme is ensured for $\theta \geq 1/2$ and therefore $\theta = 1/2$ is chosen. For a time step size $\tau$, i.e. $t^{j+1} = t^j + \tau$, the complete discretization is obtained as
	\begin{equation}\begin{aligned}
		\label{eq:STRAHL_complete_discretization}
		\bar{n}_i^{j+1} &+ \bar{n}_i^j = \tau \bar{S}_i \\
			 + \tau &\left\{ \tilde{D}_{i-} +\left[1 + K_{i-\frac{1}{2}}\right] \tilde{v}_{i-} \right\} \cdot \left( \bar{n}_{i-1}^{j+1} + \bar{n}_{i-1}^j \right) \\
			 - \tau &\left\{ \tilde{D}_{i-} -\left[1 - K_{i-\frac{1}{2}}\right] \tilde{v}_{i-} \right\} \cdot \left( \bar{n}_{i}^{j+1} + \bar{n}_{i}^j \right) \\
			 + \tau &\left\{ \tilde{D}_{i+} -\left[1 + K_{i+\frac{1}{2}}\right] \tilde{v}_{i+}  \right\} \cdot \left( \bar{n}_{i}^{j+1} + \bar{n}_{i}^j \right) \\
			 + \tau &\left\{ \tilde{D}_{i+} -\left[1 - K_{i+\frac{1}{2}}\right] \tilde{v}_{i+} \right\} \cdot \left( \bar{n}_{i+1}^{j+1} + \bar{n}_{i+1}^j \right) ~.
	\end{aligned}\end{equation}
Depending on the P\'eclet number $\mu_i = v_i \Delta r_i/D_i$, the spatial discretization is changed adaptively from a central scheme ($|\mu_i| \to 0$) to upwinding ($|\mu_i| \to \infty$) through the parameter (see figure~\ref{fig:parameter_K})
	\begin{align}
	\label{eq:STRAHL_kappa}
		K_i = \max\left(0, 1 - 2/|\mu_i|\right)\cdot \mathrm{sgn}\left(\mu_i\right) ~.
	\end{align}	
The scheme presented is applied for the discretization inside the \texttt{STRAHL} simulation domain. To treat the left and right boundary, the discretization is adjusted in accordance with the existing \texttt{STRAHL} boundary conditions.
	\ifshowfigures %
  \tikzsetnextfilename{10_parameter_K}%
  \begin{figure}[tb!]
	\centering
	\includegraphics[scale=1]{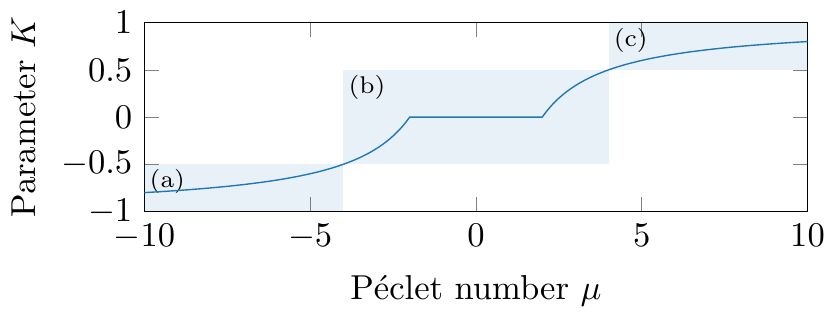}
	\vspace*{-.5cm}
	\caption[Discretization parameter $K$ dependence on P\'eclet number]{%
		\label{fig:parameter_K}
		The discretization method applied is adjusted through the parameter $K$ (see equation~\ref{eq:STRAHL_kappa}). For cases of strong advection with $|\mu| > 4$ (a,c), the backward/forward scheme dominates the discretization, whereas for strong diffusion with $|\mu| < 4$ (b), the central scheme dominates (see equation~\ref{eq:STRAHL_complete_discretization}).
		}
\end{figure}%
 \fi

\section*{References}
\addcontentsline{toc}{section}{References}
\bibliography{ms}

\begin{thebibliography}{10}

\bibitem{Reux15}
C~Reux, V~Plyusnin, B~Alper, D~Alves, B~Bazylev {\it et~al}.
\newblock Runaway electron beam generation and mitigation during disruptions at
  {JET-ILW}.
\newblock \href{http://dx.doi.org/10.1088/0029-5515/55/9/093013}{{\em Nucl.
  Fusion}} {\bf 55}, 093013 (2015).

\bibitem{Matthews16}
G~F Matthews, B~Bazylev, A~Baron-Wiechec, J~Coenen, A~Heinola {\it et~al}.
\newblock Melt damage to the {JET} {ITER}-like {W}all and divertor.
\newblock \href{http://dx.doi.org/10.1088/0031-8949/T167/1/014070}{{\em Phys.
  Scr.}} {\bf T167}, 014070 (2016).

\bibitem{Hender07}
T~C Hender, J~C Wesley, J~Bialek, A~Bondeson, A~H Boozer {\it et~al}.
\newblock Chapter 3: {MHD} stability, operational limits and disruptions.
\newblock \href{http://dx.doi.org/10.1088/0029-5515/47/6/S03}{{\em Nucl.
  Fusion}} {\bf 47}, S128 (2007).

\bibitem{Pautasso17}
G~Pautasso, M~Bernert, M~Dibon, B~Duval, R~Dux {\it et~al}.
\newblock Disruption mitigation by injection of small quantities of noble gas
  in {ASDEX} {U}pgrade.
\newblock \href{http://dx.doi.org/10.1088/0741-3335/59/1/014046}{{\em Plasma
  Phys. Control. Fusion}} {\bf 59}, 014046 (2017).

\bibitem{Coda19}
S~Coda, M~Agostini, R~Albanese, S~Alberti, E~Alessi {\it et~al}.
\newblock Physics research on the {TCV} tokamak facility: from conventional to
  alternative scenarios and beyond.
\newblock \href{http://dx.doi.org/10.1088/1741-4326/ab25cb}{{\em Nucl. Fusion}}
  {\bf 59}, 112023 (2019).

\bibitem{Commaux10}
N~Commaux, L~R Baylor, T~C Jernigan, E~M Hollmann, P~B Parks {\it et~al}.
\newblock Demonstration of rapid shutdown using large shattered deuterium
  pellet injection in {DIII}-{D}.
\newblock \href{http://dx.doi.org/10.1088/0029-5515/50/11/112001}{{\em Nucl.
  Fusion}} {\bf 50}, 112001 (2010).

\bibitem{Hesslow18}
L~Hesslow, O~Embr\'eus, G~J Wilkie, G~Papp and T~F\"ul\"op.
\newblock Effect of partially ionized impurities and radiation on the effective
  critical electric field for runaway generation.
\newblock \href{http://dx.doi.org/10.1088/1361-6587/aac33e}{{\em Plasma Phys.
  Control. Fusion}} {\bf 60}, 074010 (2018).

\bibitem{Hesslow18_2}
L~Hesslow, O~Embr\'eus, M~Hoppe, T~C DuBois, G~Papp {\it et~al}.
\newblock Generalized collision operator for fast electrons interacting with
  partially ionized impurities.
\newblock \href{http://dx.doi.org/10.1017/S0022377818001113}{{\em J. Plasma
  Phys.}} {\bf 84}, 905840605 (2018).

\bibitem{Hesslow19}
L~Hesslow, O~Embr\'eus, O~Vallhagen and T~F\"ul\"op.
\newblock Influence of massive material injection on avalanche runaway
  generation during tokamak disruptions.
\newblock \href{http://dx.doi.org/10.1088/1741-4326/ab26c2}{{\em Nucl. Fusion}}
  {\bf 59}, 084004 (2019).

\bibitem{Konovalov14}
S~Konovalov, P~Aleynikov, K~Aleynikova, Yu~Gribov, G~T~A Huijsmans {\it et~al}.
\newblock Integrated modelling of {ITER} disruption mitigation.
\newblock Contribution at the
  \href{http://www-naweb.iaea.org/napc/physics/FEC/FEC2014/fec2014-preprints/561_THP331.pdf}{25$^{\rm
  th}$ IAEA Fusion Energy Conference} 13.-18.10.2014, St. Petersburg, Russian
  Federation, TH/P3-31.

\bibitem{Fable13}
E~Fable, C~Angioni, A~A Ivanov, K~Lackner, O~Maj {\it et~al}.
\newblock Dynamical coupling between magnetic equilibrium and transport in
  tokamak scenario modelling, with application to current ramps.
\newblock \href{http://dx.doi.org/10.1088/0741-3335/55/7/074007}{{\em Plasma
  Phys. Control. Fusion}} {\bf 55}, 074007 (2013).

\bibitem{Dux99}
R~Dux, A~G Peeters, A~Gude, A~Kallenbach, R~Neu {\it et~al}.
\newblock {$Z$} dependence of the core impurity transport in {ASDEX} {U}pgrade
  {H} mode discharges.
\newblock \href{http://dx.doi.org/10.1088/0029-5515/39/11/302}{{\em Nucl.
  Fusion}} {\bf 39}, 1509 (1999).

\bibitem{Fable16}
E~Fable, G~Pautasso, M~Lehnen, R~Dux, M~Bernert {\it et~al}.
\newblock Transport simulations of the pre-thermal-quench phase in {ASDEX}
  {U}pgrade massive gas injection experiments.
\newblock \href{http://dx.doi.org/10.1088/0029-5515/56/2/026012}{{\em Nucl.
  Fusion}} {\bf 56}, 026012 (2016).

\bibitem{Putvinski97}
S~Putvinski, N~Fujisawa, D~Post, N~Putvinskaya, M~N Rosenbluth {\it et~al}.
\newblock Impurity fueling to terminate {T}okamak discharges.
\newblock \href{http://dx.doi.org/10.1016/S0022-3115(97)80056-6}{{\em J. Nucl.
  Mater.}} {\bf 241-243}, 316 (1997).

\bibitem{Feher11}
T~Feh\'er, H~M Smith, T~F\"ul\"op and K~G\'al.
\newblock Simulation of runaway electron generation during plasma shutdown by
  impurity injection in {ITER}.
\newblock \href{http://dx.doi.org/10.1088/0741-3335/53/3/035014}{{\em Plasma
  Phys. Control. Fusion}} {\bf 53}, 035014 (2011).

\bibitem{Ivanov05}
A~A Ivanov, R~R Khayrutdinov, S~Yu Medvedev and Yu~Yu Poshekhonov.
\newblock New adaptive grid plasma evolution code {SPIDER}.
\newblock Contribution at the
  \href{http://epsppd.epfl.ch/Tarragona/pdf/P5_063.pdf}{32$^{\rm nd}$ EPS
  Conference on Plasma Physics} 27.06-01.07.2005, Tarragona, Spain, P5.063.

\bibitem{Summers04}
H~P Summers.
\newblock The {ADAS} {U}ser {M}anual, version 2.6.
\newblock \href{http://www.adas.ac.uk}{http://www.adas.ac.uk}.

\bibitem{Peeters00}
A~G Peeters.
\newblock Reduced charge state equations that describe {P}firsch {S}chl\"uter
  impurity transport in tokamak plasma.
\newblock \href{http://dx.doi.org/10.1063/1.873812}{{\em Phys. Plasmas}} {\bf
  7}, 268 (2000).

\bibitem{Angioni17}
C~Angioni, R~Bilato, F~J Casson, E~Fable, P~Mantica {\it et~al}.
\newblock Gyrokinetic study of turbulent convection of heavy impurities in
  tokamak plasmas at comparable ion and electron heat flux.
\newblock \href{http://dx.doi.org/10.1088/0029-5515/57/2/022009}{{\em Nucl.
  Fusion}} {\bf 57}, 022009 (2017).

\bibitem{Hundsdorfer03}
W~Hundsdorfer and J~G Verwer.
\newblock {\em Numerical Solution of Time-Dependent
  Advection-Diffusion-Reaction Equations}.
\newblock \href{http://dx.doi.org/10.1007/978-3-662-09017-6}{Springer-Verlag},
  Berlin, 1st edition, 2003.

\bibitem{Chiu98}
S~C Chiu, M~N Rosenbluth, R~W Harvey and V~S Chan.
\newblock Fokker-{P}lanck simulations mylb of knock-on electron runaway
  avalanche and bursts in tokamaks.
\newblock \href{http://dx.doi.org/10.1088/0029-5515/38/11/309}{{\em Nucl.
  Fusion}} {\bf 38}, 1711 (1998).

\bibitem{Harvey00}
R~W Harvey, V~S Chan, S~C Chiu, T~E Evans, M~N Rosenbluth {\it et~al}.
\newblock Runaway electron production in {DIII}-{D} killer pellet experiments,
  calculated with the {CQL3D/KPRAD} model.
\newblock \href{http://dx.doi.org/10.1063/1.1312816}{{\em Phys. Plasmas}} {\bf
  7}, 4590 (2000).

\bibitem{Papp15}
G~Papp, A~Stahl, M~Drevlak, T~F\"ul\"op, Ph~W Lauber {\it et~al}.
\newblock Towards self-consistent runaway electron modeling.
\newblock Contribution at the
  \href{http://ocs.ciemat.es/EPS2015PAP/pdf/P1.173.pdf}{42$^{\mathrm{nd}}$ EPS
  Conference on Plasma Physics} 22-26.06.2015, Lisbon, Portugal, P1.173.

\bibitem{Connor75}
J~W Connor and R~J Hastie.
\newblock Relativistic limitations of runaway electrons.
\newblock \href{http://dx.doi.org/10.1088/0029-5515/15/3/007}{{\em Nucl.
  Fusion}} {\bf 16}, 415 (1975).

\bibitem{Rosenbluth97}
M~N Rosenbluth and S~V Putvinski.
\newblock Theory for avalanche of runaway electrons in tokamaks.
\newblock \href{http://dx.doi.org/10.1088/0029-5515/37/10/I03}{{\em Nucl.
  Fusion}} {\bf 37}, 1355 (1997).

\bibitem{Stahl16}
A~Stahl, O~Embr\'eus, G~Papp, M~Landreman and T~F\"ul\"op.
\newblock Kinetic modelling of runaway electrons in dynamic scenarios.
\newblock \href{http://dx.doi.org/10.1088/0029-5515/56/11/112009}{{\em Nucl.
  Fusion}} {\bf 56}, 112009 (2016).

\bibitem{Hesslow19_2}
L~Hesslow, L~Unnerfelt, O~Vallhagen, O~Embr\'eus, M~Hoppe {\it et~al}.
\newblock Evaluation of the {D}reicer runaway growth rate in the presence of
  high-{$Z$} impurities using a neural network.
\newblock \href{http://dx.doi.org/10.1017/S0022377819000874}{{\em J. Plasma
  Phys}} {\bf 85}, 475850601 (2019).

\bibitem{Papp13}
G~Papp, T~F\"ul\"op, T~Feh\'er, P~C de~Vries, V~Riccardo {\it et~al}.
\newblock The effect of {ITER}-like wall on runaway electron generation in
  {JET}.
\newblock \href{http://dx.doi.org/10.1088/0029-5515/53/12/123017}{{\em Nucl.
  Fusion}} {\bf 53}, 123017 (2013).

\bibitem{Pautasso07}
G~Pautasso, C~J Fuchs, O~Gruber, C~F Maggi, M~Maraschek {\it et~al}.
\newblock Plasma shut-down with fast impurity puff on {ASDEX} {U}pgrade.
\newblock \href{http://dx.doi.org/10.1088/0029-5515/47/8/023}{{\em Nucl.
  Fusion}} {\bf 47}, 900 (2007).

\bibitem{Fil15}
A~Fil.
\newblock {\em Modeling of massive gas injection triggered disruptions in
  tokamak plasmas}.
\newblock PhD thesis,
  \href{https://www.theses.fr/en/2015AIXM4040}{Aix-Marseille University},
  September 2015.

\bibitem{Igochine10}
V~Igochine, A~Gude, M~Maraschek and the ASDEX Upgrade~Team.
\newblock {\em Hotlink based {S}oft {X}-ray {D}iagnostic on {ASDEX} {U}pgrade}.
\newblock Number IPP 1/338.
  \href{http://hdl.handle.net/11858/00-001M-0000-0026-F036-B}{Max-Planck-Institut
  f\"ur Plasmaphysik}, May 2010.

\end{thebibliography}

\end{document}